\documentclass{article}

\usepackage{arxiv}
\usepackage[utf8]{inputenc} 
\usepackage[T1]{fontenc}    
\usepackage{hyperref}       
\usepackage{url}            
\usepackage{booktabs}       
\usepackage{amsfonts}       
\usepackage[numbers]{natbib}
\usepackage{amsmath, enumitem}
\usepackage{amssymb}
\usepackage{upgreek}
\usepackage{lineno}
\usepackage{soul}
\usepackage{fixltx2e}

\usepackage{nccmath}
\usepackage{ifthen}
\usepackage{color}
\usepackage{graphicx}
\usepackage{algorithm}
\usepackage{algorithmic}
\usepackage{setspace}
\usepackage{multirow}
\usepackage[T1]{fontenc}

\DeclareMathOperator*{\argmax}{argmax}
\graphicspath{{Fig/}}

\title{A Semi-analytic but Biased Uncertainty Assessment Method using Sample Extensions, Analysed for Nonlinear\\ Travel Time Tomography}
\date{} 					

\author{
	Xuebin Zhao \\
	School of Geosciences \\
	University of Edinburgh\\
	Edinburgh, Unite Kingdom \\
	\And
	Andrew Curtis \\
	School of Geosciences \\
	University of Edinburgh\\
	Edinburgh, Unite Kingdom \\
}

\begin{document}
	\maketitle

\begin{abstract}
Many geophysical problems can be cast as inverse problems that estimate a set of parameter values from observed data. Within a Bayesian framework, solutions to such problems are described probabilistically by the so-called \textit{posterior} probability distribution functions (pdf's) which combines \textit{a priori} information with information from measured data. To obtain robust inference results often requires millions of model parameter value samples to be drawn, and simulation of the dataset that would have been recorded if each of these values was true; this is a computationally expensive procedure. We investigate the concept of \textit{sample extensions} as a means to improve efficiency when solving fully nonlinear inverse problems. A sample's extension is defined as the set of models or parameter values whose corresponding forward function values are directly accessible from a sample for which the forward function has already been evaluated, obviating the need for additional forward function evaluations. In a specific case of ray-based first-arrival travel time calculations which are often used in seismic travel time tomography, we apply sample extensions to obtain a continuous region with non-zero hypervolume within parameter space, across all of which the forward function values are known given only a single forward simulation. We devise a deterministic sampling technique that identifies the most informative extensions by solving an optimisation problem. In an illustrative tomographic example that involves a single travel time datum, we find 51 optimal samples, and use them to construct an \textit{analytic} approximation to the Bayesian posterior pdf. Additionally, we propose an extensions-based algorithm for real-world tomography scenarios and apply it to a synthetic 2D example. This study highlights two fundamental problems that make the method inefficient: (1) limited hypervolumes of extensions and (2) neglecting parameter correlations to simplify analytic calculations. Finding solutions to these problems defines possible directions for future research.
\end{abstract}

\section{Introduction}
Geophysical imaging is used to visualise the Earth's interior by estimating the subsurface heterogeneity of properties, such as seismic velocity or resistivity values. This involves solving an inverse problem to estimate model parameter values that describe the intrinsic nature of the Earth given data acquired on the surface. This often requires a so-called \textit{forward} function $\boldsymbol f$ to be defined, which maps the space of parameter values (henceforth, \textit{model} space) to the data space by simulating synthetic data that would be observed under the assumption that the true Earth corresponds to a specific set of parameter values. Subsequently, we invert this mapping to estimate the set of parameter solutions that are consistent with the observed data \cite{valentine2023emerging, tsai2023future}.

A common objective is to find the globally optimal solution that best fits the recorded data. While this may appear straightforward, in cases where the forward function is nonlinear an algorithm that searches model parameter space may fail to find that solution, and in cases in which the data are contaminated by noise, the best fit solution may anyway not be the best approximation to the true Earth \cite{tarantola2005inverse}. 

We might instead seek to estimate the family of all possible parameter values that are consistent with the observed data and their uncertainties. Commonly this family is referred to as the parameter uncertainty \cite{tarantola2005inverse}. Methods to estimate the uncertainty often solve inverse problems within a probabilistic framework, by calculating statistics of the \textit{posterior} probability distribution function (pdf) constructed using Bayes rule. This combines \textit{prior} knowledge about parameter values with information from the observed data, an approach known as \textit{Bayesian inference}. Uncertainty information in the inversion results can in theory then be quantified. However, if we are to be sure that all possible sets of parameter values have been assessed in order to provide confidence in the result, in principle both optimisation-based and probability-based assessment methods would require the entire parameter space to be explored, and a misfit between the observed and predicted data to be evaluated at each point in that space. This is impossible.

Sampling-based methods explore parameter space using differently justified algorithms to identify model parameter values that are consistent with the observed data. Various techniques have been proposed to generate deterministically gridded samples \cite[e.g.,][]{lomax2001fast, lomax2009earthquake}, or pseudo-random samples from algorithms that include Monte Carlo methods \cite{press1968earth, mosegaard1995monte, malinverno2002parsimonious}, simulated annealing \cite{kirkpatrick1983optimization, sen2013global}, genetic algorithms \cite{stoffa1991nonlinear, sambridge1992genetic, sambridge1993earthquake}, the neighbourhood algorithm \cite{sambridge1999geophysical, sambridge1999geophysical2}, and prior sampling \cite{kaufl2016solving, mosser2020stochastic, bloem2022introducing}. The \textit{curse of dimensionality} states that the number of samples required to explore parameter space with a desired sample density grows exponentially with the dimensionality of the problem \cite{curtis2001prior}. Therefore, while these approaches are effective they are often computationally expensive to implement, especially when the problem's dimensionality (the number of unknown parameters to be inferred) is high. 

Effective strategies have been introduced to geophysics to improve the efficiency of these sampling-based methods. Trans-dimensional algorithms reduce the dimensionality to only those parameters that are essential to explain the data \cite{green1995reversible}. This reduction diminishes the required number of samples significantly \cite{bodin2009seismic, bodin2012transdimensionaltomo, galetti2017transdimensional}; however the trans-dimensional nature of the algorithm calls into question whether the results are physically meaningful, because Bayes theorem does not give consistent answers when different parametrisations of the model are used -- even if those parametrisations are in principle exactly equivalent \cite{mosegaard2024inconsistency}. Techniques that incorporate gradient information from the logarithmic posterior probability have also efficiently located high probability regions compared to pure Monte Carlo methods \cite{fichtner2018hamiltonian, fichtner2019hamiltonian, gebraad2020bayesian, zhao2021gradient, de2023acoustic}. Other studies have concentrated on exact sampling methods, which ensure that every sample is exactly a sample of the posterior pdf; thus, sampling is unbiased so that samples contain maximum information \cite{propp1996exact, walker2014spatial}, with analytic (non-sampling-based) versions of these algorithms available in some specific classes of problems \cite{nawaz2018variational}. Nevertheless, most sampling-based methods (except for exact sampling methods) are slow to converge, and all sampling algorithms are only proven to converge in infinite time rendering interpretations of any finite-time sample sets potentially biased \cite{atchade2005adaptive, andrieu2008tutorial}. In addition, detecting when convergence to a stationary distribution has occurred so that results can be interpreted with confidence remains a challenge in practical problems \cite{besag1993spatial, walker2014spatial}.

Another class of methods, \textit{variational inference}, has gained widespread application across various Bayesian geophysical problems \cite{rezende2015variational, liu2016stein, guo2016boosting, gallego2018stochastic, nawaz2018variational, nawaz2019rapid, zhang2019seismic, zhao2021bayesian, siahkoohi2021preconditioned, smith2022hyposvi, levy2022variational, wang2023re, lomas20233d, zhao2024physically, zhao2024efficient}. The objective of variational methods is to identify an optimal distribution from a known family of distributions, which best approximates the true posterior pdf. A sub-category of essentially variational methods leverages neural networks (NNs) to solve inverse problems. By employing a substantial collection of samples probabilistically distributed based on prior knowledge and their corresponding modelled data, NNs are trained to approximate a nonlinear mapping from data space to model parameter space using optimisation. After training, these networks enable fast inversion of any new data sets to obtain model parameter values and their associated uncertainties, which can be achieved by a forward pass through the trained NNs \cite{devilee1999efficient, meier2007fully, meier2007global, ray2010efficient, shahraeeni2011fast, shahraeeni2012fast, de2013bayesian, kaufl2014framework, laloy2018training, earp2019probabilistic, earp2019probabilistic2, cao2020near, lubo2021exhaustive, zhang2021bayesian, hansen2022use, levy2022using, bloem2022introducing}. Both variational methods and NNs are optimisation-based methods, which confers an advantage over Monte Carlo methods in terms of detecting convergence \cite{goodfellow2016deep}.

A common feature of all of the methods described above is that the improved efficiency is obtained by advanced mathematical formulation of the inverse problems. However, the \textit{No Free Lunch theorem} states that no algorithm outperforms any other when evaluated across all problems \cite{wolpert1997no}. This implies that any method might be the most effective for a certain class of problems, but not for all problems. Contrapositively, this theorem also implies that there is merit in the search for algorithms that best suit particular (perhaps narrow) classes of problems. 

We can define classes of problems that have a common forward function (hence, common physics relating model parameters and data). In all of the algorithms cited above, forward function values are obtained only for parameter values set by each model sample. In volumes of parameter space that remain unsampled, the forward function values remain unknown. Since the main computational expense of sampling algorithms tends to occur in forward function evaluation, we conclude that as much information as possible should be extracted from each such evaluation.

In many geophysical problems we understand physical characteristics of the forward functions beyond their numerical evaluation at a point. In most instances, the continuity and smoothness of forward functions are considered when designing sampling algorithms, while other properties are disregarded. Yet other untapped knowledge has the potential to enhance the efficiency of parameter space exploration. In this work we aim to create algorithms that use this knowledge effectively.

\citet{curtis2020samples} analysed symmetries intrinsic to the forward functions of certain problem classes and showed that additional information is available from each model sample $\mathbf{m}$ at almost no additional computational cost. This extra information was referred to as the \textit{extension} of sample $\mathbf{m}$, defined as all information about other forward function evaluations that is available from $\boldsymbol{f} (\mathbf{m})$ without performing any further explicit evaluations of $\boldsymbol{f}$. In common geophysical problems it was shown that for some forward functions, the extension spans continuous hypervolumes of parameter space, an infinite increase in information compared to the original forward function value. More importantly, as the number of parameters and hence model dimensionality increases, the information provided by these extensions provides an almost proportionate increase in information. This suggests that there may be potential to alleviate challenges imposed by the curse of dimensionality.

This study investigates a specific application of sample extensions to travel time tomographic problems. In Section 2, we establish theoretical foundations, including two types of extensions that are specific to the forward problem commonly considered in travel time tomography problems. Then we propose a deterministic sampling algorithm designed to find the most informative samples and their corresponding extensions. In Section 3, we present an extensions-based algorithm for travel time tomography. Two examples are provided in the subsequent section: the first shows the effectiveness of the proposed tomography algorithm, and the second serves to pinpoint the primary sources of errors inherent to the algorithm. This paves the way for potential directions in future research. Finally, we discuss the implications of our findings and draw conclusions.

\section{Sample Extensions}
\subsection{Fundamentals}
\begin{figure}
	\centering\includegraphics[width=0.4\textwidth]{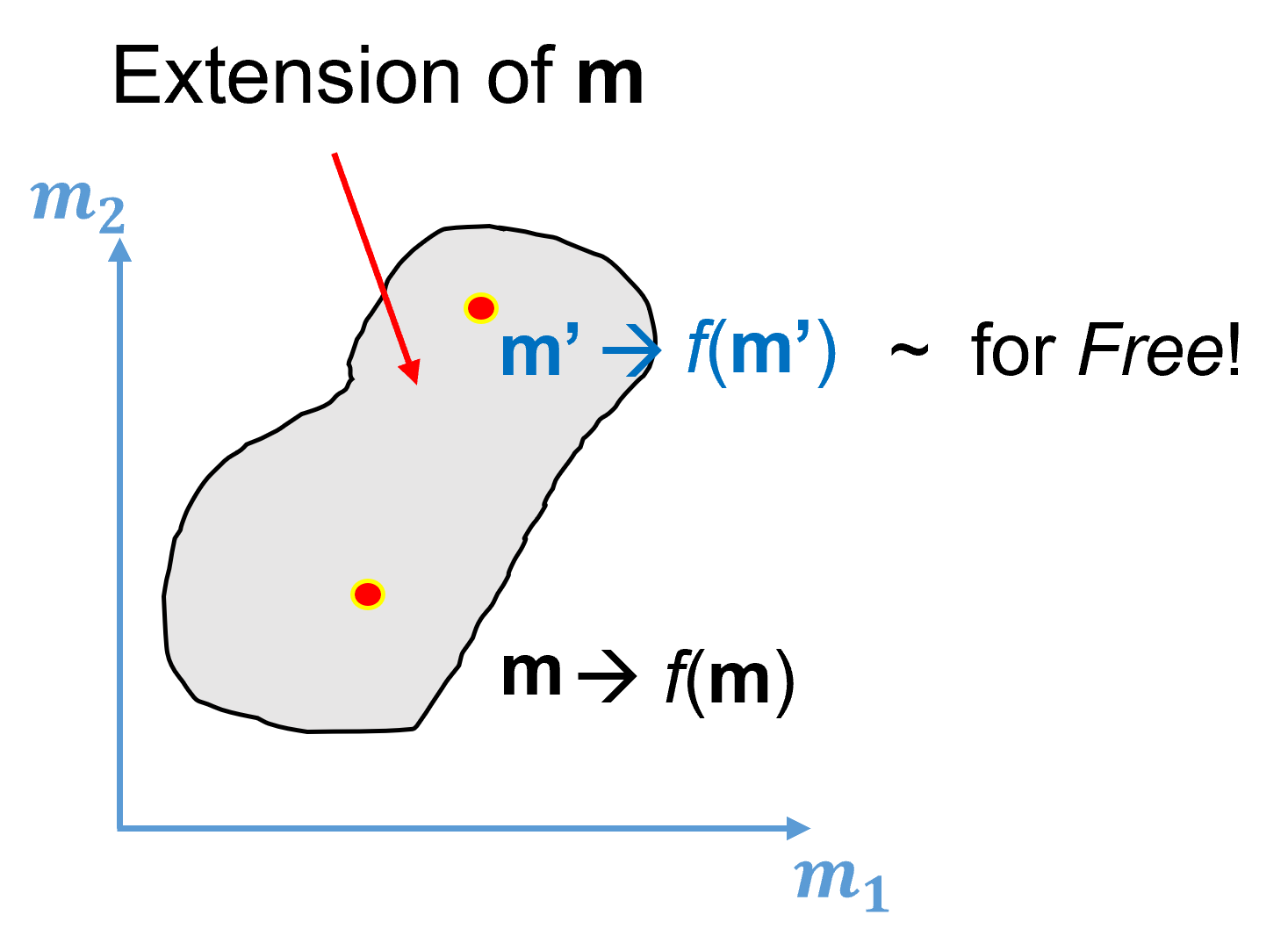}
	\caption{General concept of sample extensions in a 2-dimensional parameter space. We select a sample $\mathbf{m}$ (lower red point) and calculate its forward function value $\boldsymbol f(\mathbf{m})$. Leveraging the inherent physical properties of the forward function $\boldsymbol f(\cdot)$ and definition of parameters in $\mathbf{m}$, we obtain forward function values $\boldsymbol f(\mathbf{m^\prime})$ for other model samples $\mathbf{m^\prime}$ immediately within the grey area without additional evaluations of $\boldsymbol f$. The subset of parameter space defined by all such possible samples is called the \textit{extension} of sample $\mathbf{m}$.}
	\label{fig:extension_2d}
\end{figure}

The concept of \textit{sample extensions} was introduced originally by \citet{curtis2020samples}. Consider a known forward function value $\boldsymbol f(\mathbf{m})$ corresponding to a specific sample $\mathbf{m}$. The extension of $\mathbf{m}$ is defined as a set of forward function values for other model parameter values $\mathbf{m^\prime}$ that can be derived from the existing model sample $\mathbf{m}$ without further forward function evaluations. This concept is illustrated in Figure \ref{fig:extension_2d}: if we have already calculated the forward function value $\boldsymbol f(\mathbf{m})$ for a random sample $\mathbf{m}$ in a two-dimensional parameter space, we can obtain function values $\boldsymbol f(\mathbf{m^\prime})$ for other samples $\mathbf{m^\prime}$ immediately based on our understanding of some physical properties of the forward function $\boldsymbol f(\cdot)$ and definitions of parameters $\mathbf{m}$, rather than by conducting additional forward simulations. The set of all such samples $\mathbf{m^\prime}$ illustrated by the grey region in Figure \ref{fig:extension_2d} is called the \textit{extension} of $\mathbf{m}$.

Extensions offer the possibility to explore parameter space and solve inverse problems more cheaply, primarily due to their capacity to provide information about the forward function of parameter $\mathbf{m^\prime}$ at a considerably reduced cost compared to explicit forward evaluations. Types of extensions tend to differ across inverse problems, due to differences in the underlying physical relationships in $\boldsymbol f$, and a more comprehensive overview of extensions can be found in \citet{curtis2020samples}. In this paper, we focus on a specific inverse problem type, by showing how extensions can be used to solve fully nonlinear travel time tomography problems.

Seismic travel time tomography is an inverse problem where subsurface slowness (or velocity) map of a spatial domain through which wave energy has traversed is estimated, by using observed travel times of waves between source and receiver locations. In seismic tomography problem, the forward function $\boldsymbol f(\mathbf{m})$ predicts the travel times of first arriving waves propagating through any given model $\mathbf{m}$ which describes the spatial distribution of wave speed, velocity or slowness (the reciprocal of velocity). Usually this function is computed by solving the eikonal equation using the fast marching method \cite[FMM --][]{rawlinson2004wave, rawlinson2005fast}. 

We introduce two types of extensions based on characteristics of this forward function:
\begin{itemize}
	\item \textbf{Off-ray extension}: Increasing off-ray model slowness values does not modify the trajectory of the fastest ray path or the corresponding first arrival travel time.
	\item \textbf{On-ray extension}: Decreasing on-ray model slowness values does not alter the trajectory of the fastest ray path, and the corresponding travel time is predictable.
\end{itemize}

\begin{figure}
	\centering\includegraphics[width=\textwidth]{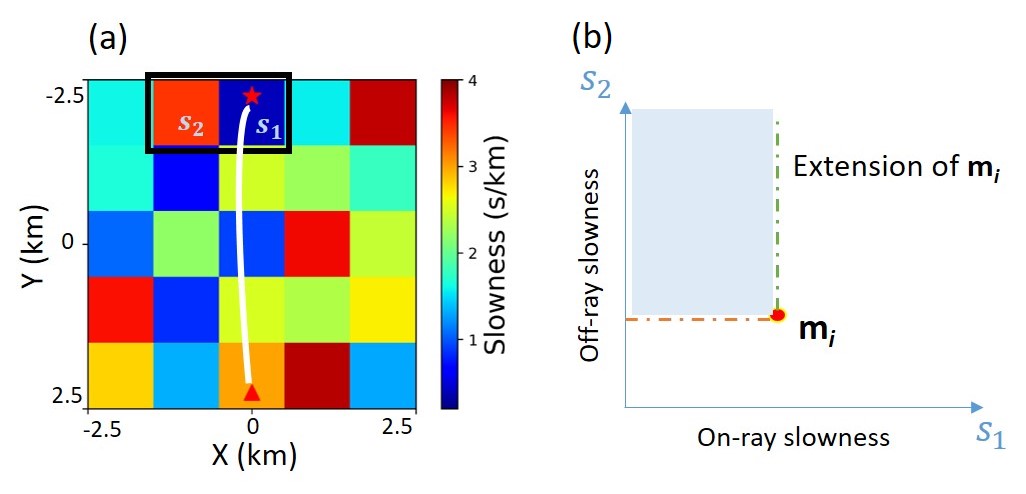}
	\caption{Extensions for travel time prediction. (a) A random slowness sample $\mathbf{m}_i$ with a known forward function value $f(\mathbf{m}_i)$. Source and receiver locations are denoted by red star and triangle, respectively. The white curve represents schematically the first arriving ray path between them. (b) Parameter space for two slowness cells $S_1$ and $S_2$, located inside the black box in (a). Point $\mathbf{m}_i$ represents the slowness model sample in (a), and the blue region shows the extensions subspace of this sample. By employing the on-ray and off-ray extensions along cells $S_1$ and $S_2$, respectively, we encompass a parameter subspace within which forward function values are known based on the already-calculated value of $\boldsymbol f(\mathbf{m}_i)$.}
	\label{fig:extension_illustration}
\end{figure}

Figure \ref{fig:extension_illustration}a displays a random slowness model sample defined over a square-cellular grid, and (schematically) the corresponding first arrival ray path (white curve) between one source (red star) and one receiver (red triangle). Ray theory states that if we increase the slowness of cells that are not transversed by the ray -- referred to as off-ray cells, the fastest ray remains the same and, so does the corresponding travel time \cite{curtis2020samples}. This is illustrated in Figure \ref{fig:extension_illustration}b. Considering an off-ray cell $S_2$ inside the black box in Figure \ref{fig:extension_illustration}a, the \textit{off-ray extension} permits a limitless increase in its slowness value with no effect on the ray path or travel time. Consequently, forward function values for all model samples positioned along the dashed green line in Figure \ref{fig:extension_illustration}b remain identical and hence known.

We also consider an alternative physical approximation wherein the first arrival energy travelling along a ray propagates through forward scattering or diffraction by each grid cell, and hence obeys scattering theory rather than strictly following Snell's law. This is valid for infinitely small cells, and is approximately true for finite-sized but small cells. From this perspective, slownesses associated with cells along the trajectory of the ray may be decreased without changing the fastest ray path; the new travel time value can be calculated by adding the original travel time to the product of the difference in slowness values and the length of the ray within each cell. We call this the \textit{on-ray extension}. A more comprehensive discussion of the on-ray extension is presented in Appendix \ref{ap:A}. In Figure \ref{fig:extension_illustration}b, we show one on-ray cell $S_1$ and its extensions. By applying the on-ray extension, we know forward function values for models on the dashed orange line almost for free. Ultimately, if we apply the on-ray extension for all models within the off-ray extension of $\mathbf{m}_i$ -- models positioned along the dashed green line in Figure \ref{fig:extension_illustration}b -- we effectively obtain a 2-dimensional subspace (the blue region in Figure \ref{fig:extension_illustration}b) inside which the forward function values (travel times) are known at all points, given only the forward function evaluation at the single point $\mathbf{m}_i$. Where the context makes this unambiguous, we refer to the combination of the off-ray and on-ray extensions simply as the extensions of $\mathbf{m}_i$.

This indicates that we can apply either the on-ray or off-ray extension for any slowness cell in Figure \ref{fig:extension_illustration}a. Consequently, for any slowness model sample, we obtain a non-zero extension hypervolume within the full parameter space, and with only one forward evaluation we gain immediate access to forward function values for an infinite number of other model samples. The potential value of this information within tomography problems has hitherto remained unexplored.

\subsection{Deterministic Selection of Extensions}
\citet{curtis2020samples} presented a 3-parameter toy tomography example that uses extensions for travel time prediction. Fifteen random samples were drawn within parameter space, and the resulting extensions of these samples yield continuous, volumetric information about the forward function values. Without using extensions, these samples offer no continuous information across the parameter space, leaving open the possibility that high probability regions lie between samples and so remain undetected. This example proved the efficacy of extensions for exploring parameter space. Of potentially greater significance, the study proceeded to select 13 non-random (deterministic) samples, each selected to have distinct ray geometries. The extensions arising from these samples provide extensive coverage of the full parameter space. The tomographic inverse problem could thus be solved to produce an accurate estimate of the posterior solution with only 13 deterministic samples (thus only 13 forward calculations). 

That example yielded two insights. First, extensions provide additional and valuable information that in principle may be used to address challenges posed by nonlinear tomography, source location, and analogous problems. Second, deterministic selection of samples may maximise information provided by extensions. In that example, this was equivalent to selecting samples that generate the largest extensions hypervolume, such that we know forward function values across as extensive a portion of the parameter space as possible given a limited budget of forward evaluations. For travel time tomography, the 3-parameter toy example in \citet{curtis2020samples} indicated that this might be achieved by sampling ray paths instead of slowness model samples. 

In our current work, we use a finite set of ray paths $\mathcal{R} = \{R_1,R_2,R_3,...,R_t\}$ to provide an approximate representation of all possible rays between two spatial locations. For any individual ray $R_i$, we define a vector of ray path lengths $\mathbf{l} = (l^1,l^2,...,l^n)$ where each element $l^j \ge 0$ denotes the length of that ray within the indexed grid cell $j$ in the discretised slowness model -- which in turn is defined as a vector $\mathbf{m} = (m_1,m_2,...,m_n)$, with each $m_j \in (m_{min}, m_{max})$ representing the slowness value in one cell. Here, $m_{min}$ and $m_{max}$ are the lower and upper bounds of the slowness values. Travel time for any ray $R_i$ with ray length vector $\mathbf{l}_i$ can be represented linearly as:
\begin{equation}
	f(\mathbf{m}; \mathbf{l}_i) = \mathbf{l}_i^T\mathbf{m}
	\label{eq:tt_cal1}
\end{equation}
Define the \textit{ray dictionary} matrix $\mathbf{L} = (\mathbf{l}_1, \mathbf{l}_2,...,\mathbf{l}_t)$ to represent (approximately) the length in each cell of all possible rays. Then the first arrival travel time of a slowness model $\mathbf{m}$ can be calculated by
\begin{equation}
	f(\mathbf{m}) = min(\mathbf{L}^T\mathbf{m})
	\label{eq:tt_cal_matrix}
\end{equation}

For any given slowness model $\mathbf{m}$ and a specific ray path $\mathbf{l}_i$, the total probability mass spanned by the extension of $\mathbf{m}$ can be calculated by
\begin{equation}
	V_{ext}(\mathbf{m}, \mathbf{l}_i) = \int_{\mathbf{m}_{min}}^{\mathbf{m}_j} p(\mathbf{m})d\mathbf{m}_j \cdot
	\int_{\mathbf{m}_k}^{\mathbf{m}_{max}} p(\mathbf{m})d\mathbf{m}_k
	\label{eq:ext_v}
\end{equation}
where indices $j \in \{j:l^j>0\}$ and $k \in \{k:l^k=0\}$ denote cells that lie on and off ray path $\mathbf{l}_i$, respectively. The set $\{\forall j, \forall k\}$ encompasses indices for all cells of a slowness model. The condition $V_{ext}(\mathbf{m}, \mathbf{l}_i) > 0$ holds provided that the extension spans non-zero hyper-volumes of the prior pdf, and that no cell slownesses satisfy $m_j = m_{min}$ nor $m_k = m_{max}$. 

In the absence of observed data, we use the prior probability distribution to calculate equation \ref{eq:ext_v}, which quantifies the portion of the prior probability mass that is spanned by the extension of a slowness model $\mathbf{m}$ for a specific ray $\mathbf{l}_i$. If we assume that the prior pdf can be evaluated freely and is a proper distribution function normalised to have total probability mass of 1, then $V_{ext}(\mathbf{m}, \mathbf{l}_i)$ measures the proportion of the prior probability density over which not only the prior pdf but also the component of the likelihood function that concerns ray $\mathbf{l}_i$ can be calculated without additional forward function evaluations.

Our aim is to use deterministic sampling to find slowness model samples that maximise the cumulative prior probability mass contained within their extensions. Specifically for path $\mathbf{l}_i$, we seek the slowness model that satisfies the following optimisation problem
\begin{equation}
	\mathbf{m} = \argmax_{\mathbf{m} \in prior} V_{ext}(\mathbf{m}, \mathbf{l}_i) \qquad s.t. \qquad
	\mathbf{l}_i^T\mathbf{m} \le \mathbf{l}^T\mathbf{m}, \quad \text{for } \forall \ \mathbf{l} \in \mathbf{L}
	\label{eq:optimiser}
\end{equation}
where the inequality constraint ensures that ray $\mathbf{l}_i$ is the fastest ray. By substituting each ray for $\mathbf{l}_i$ in equation \ref{eq:optimiser} and solving this optimisation problem for every ray across the collection of all potential rays, we derive optimal slowness model samples for all considered rays.

In a special case where the prior pdf for slowness values is a uniform distribution with $p(\mathbf{m}) = 1/(m_{max} - m_{min})^n$ where $n$ represents the dimensionality of the slowness model vector $\mathbf{m}$, equation \ref{eq:ext_v} becomes 
\begin{equation}
	V_{ext}(\mathbf{m}, \mathbf{l}_i) = \prod_{j} \dfrac{m_j-m_{min}}{m_{max} - m_{min}} \cdot 
	\prod_{k} \dfrac{m_{max} - m_k}{m_{max} - m_{min}}
	\label{eq:ext_volume_uniform}
\end{equation}
where $j$ and $k$ are defined as in equation \ref{eq:ext_v}. Equation \ref{eq:ext_volume_uniform} then simply calculates the hypervolume of the extension subspace. The optimisation problem in equation \ref{eq:optimiser} requires a combination of maximising slowness values for on-ray cells, whilst minimising those for off-ray cells, to amplify $\prod_{j} \frac{m_j-m_{min}}{m_{max} - m_{min}}$ and $\prod_{k} \frac{m_{max} - m_k}{m_{max} - m_{min}}$ respectively. In the mean time, the result must satisfy the constraint that the considered ray remains the fastest possible path. If on-ray slownesses are too large or off-ray slownesses too small, the ray path of interest could cease to be the fastest path between source to receiver locations and the extension would be associated with a different ray. 

We reformulate the optimisation problem by combining equations \ref{eq:optimiser} and \ref{eq:ext_volume_uniform} as:
\begin{equation}
	\begin{split}
		&\text{minimise}  \qquad f_0(\mathbf{m}) = -\sum_{j}\log(m_j-m_{min}) - \sum_{k}\log(m_{max} - m_k) \\
		&\text{subject to} \qquad (\mathbf{l}_i - \mathbf{l})^T \mathbf{m} \le 0,\quad m_{min} < m_j \le m_{max}, \quad m_{min} \le m_k < m_{max}\\
	\end{split}
	\label{eq:optimiser_ipm}
\end{equation}
This objective function is easy to evaluate, and has an easily calculated gradient vector and a diagonal Hessian matrix containing non-zero elements; the latter allows straightforward computation of the inverse of the Hessian matrix such that we can use the second-order derivative information of the objective function to solve the optimisation problem efficiently. For optimisation, we employ the \textit{interior-point method} to tackle this inequality-constrained problem \cite{byrd1999interior, boyd2004convex, Ribeiro2017a}. The interior-point method reformulates inequality constraints into the objective function by defining and minimising the following new objective function
\begin{equation}
	\medmath{
		f_0(\mathbf{m}) - \frac{1}{q}\left\{ \log [(\mathbf{l} - \mathbf{l}_i)^T \mathbf{m}]
		+ \log(m_{max} - m_j) + \log(m_j - m_{min})	+ \log(m_{max} - m_k) + \log(m_k - m_{min})\right\}
	}
	\label{eq:barrier_ipm}
\end{equation}
where $q>0$ is a hyper-parameter that controls the accuracy of the interior-point method and is increased iteratively during optimisation to ensure that the algorithm convergences to the true solution in equation \ref{eq:optimiser_ipm} \cite{boyd2004convex}. This approach allows us to exploit the information derived from the first and second-order derivatives of the objective function, so that we can identify the solution that minimises the original objective function (equation \ref{eq:optimiser_ipm}) while satisfying the constraints.

\section{Semi-analytic Travel Time Tomography using Extensions}
\subsection{Prior Information from Extensions}
\begin{figure}
	\centering\includegraphics[width=\textwidth]{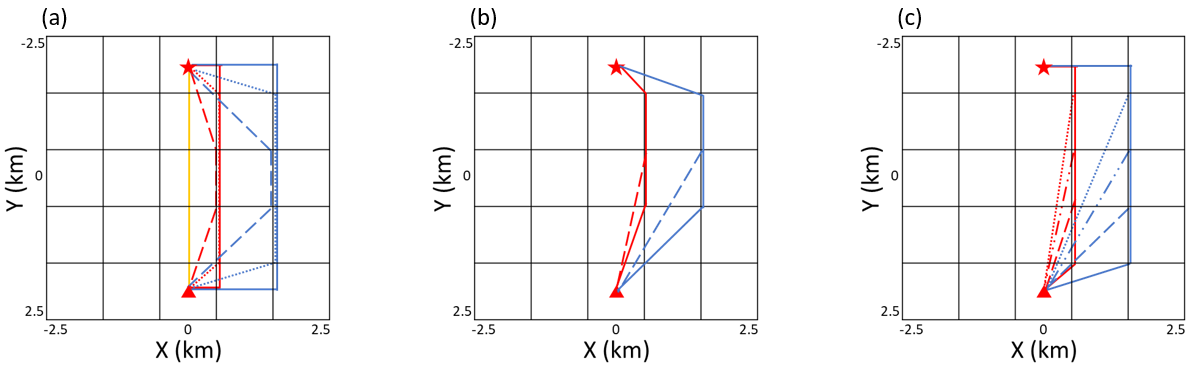}
	\caption{Grid system and ray paths used in an illustrative example. The slowness field is parametrised by a $5\times 5$ regular gridded system with a cell size of 1 km in both directions. In each figure, red star and triangle indicate source and receiver locations. The line segments that connect the two locations depict the source-receiver ray paths considered in this example. Rays in (a) are flipped horizontally around the centre line to generate 13 different rays, and rays in (b) and (c) are flipped vertically and horizontally to create 16 and 32 rays, respectively. This results in a total number of 61 rays, collectively representing an approximation to the set of all possible rays between the source and receiver.}
	\label{fig:ray_paths}
\end{figure}

We use an illustrative example to demonstrate the deterministic sampling algorithm established in the previous section. As shown in Figure \ref{fig:ray_paths}, we parametrise the slowness field into a $5 \times 5$ regular gridded system with a cell size of 1 km in both directions. Prior information for the slowness value in each cell is defined by a uniform pdf bounded between 0.2 s/km and 4.0 s/km. Locations of one source and one receiver are marked by a red star and a red triangle, respectively. Some rays are displayed as line segments that connect the source and receiver locations in Figure \ref{fig:ray_paths}. By flipping the rays in Figure \ref{fig:ray_paths}a horizontally and flipping those in Figures \ref{fig:ray_paths}b and \ref{fig:ray_paths}c vertically and horizontally around central vertical and horizontal symmetry axes, we generate 61 different rays in total. For each ray, we calculate the ray path length vector $\mathbf{l}$, and combine them together to construct the ray dictionary matrix $\mathbf{L}$ which offers an approximate representation of all feasible rays between the source and receiver locations.

We solve the optimisation problem defined in equation \ref{eq:optimiser_ipm} for each of the 61 rays. Only 51 of them yield valid slowness models because for some rays it is impossible to identify a slowness model sample within the prior space that satisfies the inequality constraints in equation \ref{eq:optimiser_ipm}. Such rays cannot be the fastest ray for any model sample within the valid range of parameter values.

\begin{figure}
	\centering\includegraphics[width=0.4\textwidth]{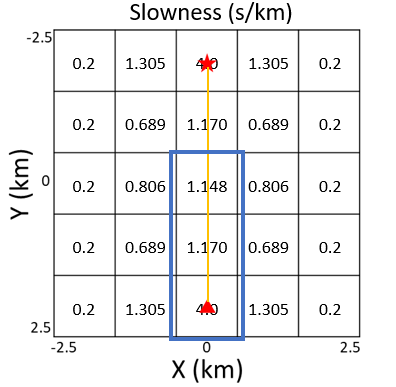}
	\caption{The optimal slowness model sample corresponding to the straight ray path (yellow line) between source (red star) and receiver (red triangle) locations, obtained using the deterministic sampling algorithm by solving the optimisation problem in equation \ref{eq:optimiser_ipm}. Number in each cell represents the slowness value for that cell.}
	\label{fig:optimal_sample_1}
\end{figure}

Figure \ref{fig:optimal_sample_1} shows one optimal slowness model corresponding to the direct ray path that connects the source and receiver locations (the yellow ray in Figure \ref{fig:ray_paths}a). We compute the extension hypervolume provided by this sample using equation \ref{eq:ext_volume_uniform}, which spans $0.175\%$ of the full prior space. This illustrates the advantage of sample extensions: with only a single forward evaluation for the model sample represented by Figure \ref{fig:optimal_sample_1}, we instantly acquire knowledge of the forward function values for all models within a continuous subspace that encompasses $0.175\%$ of the full 25-dimensional parameter space.

This may seem like a small percentage, but one should remember the vastness of 25-dimensional space. Sampling such a space with a grid of even only two parameter values per dimension requires $2^{25}$ or 33 million samples. Each of these would require a forward function evaluation, after which still only extremely sparse sampling is achieved. On the other hand, the above sample extension already spans almost one $500^{th}$ of the space continuously with only a single forward function evaluation. 

To explore the value of optimising the selection of samples, we estimate the expected extension hypervolume from a random Monte Carlo sample by drawing 1 million random samples from the prior distribution, calculating the percentage hypervolume of the extension of each sample using equation \ref{eq:ext_volume_uniform}, and taking the average across all samples. The result is $6.14\times 10^{-8} \%$, approximately 6 orders of magnitude smaller than the volume offered by the optimal deterministic sample. Finally, note that without sample extensions we merely obtain a forward function value for a solitary point which represents a hypervolume of zero within a 25-dimensional space. 

These three approaches (deterministic optimisation, random sampling, and the use of single points or models) underline the effectiveness of extensions in exploring forward function values across parameter space. The deterministic sampling approach enhances this effectiveness by maximising the information obtained from each single forward function evaluation and its extension. By applying deterministic sampling to maximise the extensions of all 51 samples above, the total volume of their extensions span over $1.1\%$ of the full prior parameter space. Equivalently, an average of $83.5\%$ of each parameter axis is spanned piecewise continuously (since $1.1\% \approx 0.835^{25}$). 

Another important aspect of the optimal sample displayed in Figure \ref{fig:optimal_sample_1} is that travel time calculated along the yellow path in Figure \ref{fig:optimal_sample_1} is equal to those calculated along 10 additional rays displayed in Figure \ref{fig:ray_paths}, which indicates that this sample has a multi-pathed first arrival; in other words, when the source is fired, the first arriving energy arrives simultaneously along all 11 rays. We can therefore apply both the on-ray and off-ray extensions to each of these 11 rays, given that the different ray geometries define different on-ray and off-ray cells and thus different extensions, inside all of which we know forward function values almost for free. This means that the total extensions hypervolume becomes even larger when accounting for multi-pathing (since we obtain 10 additional extensions). If we include multi-pathing for all 51 optimal samples, we obtain 709 distinct extensions, which in total span $3.3\%$ or one thirtieth of the full prior parameter space (corresponding to an average of $87.3\%$ per parameter axis); this information is obtained from the forward evaluations of only 51 models.

\begin{figure}
	\centering\includegraphics[width=\textwidth]{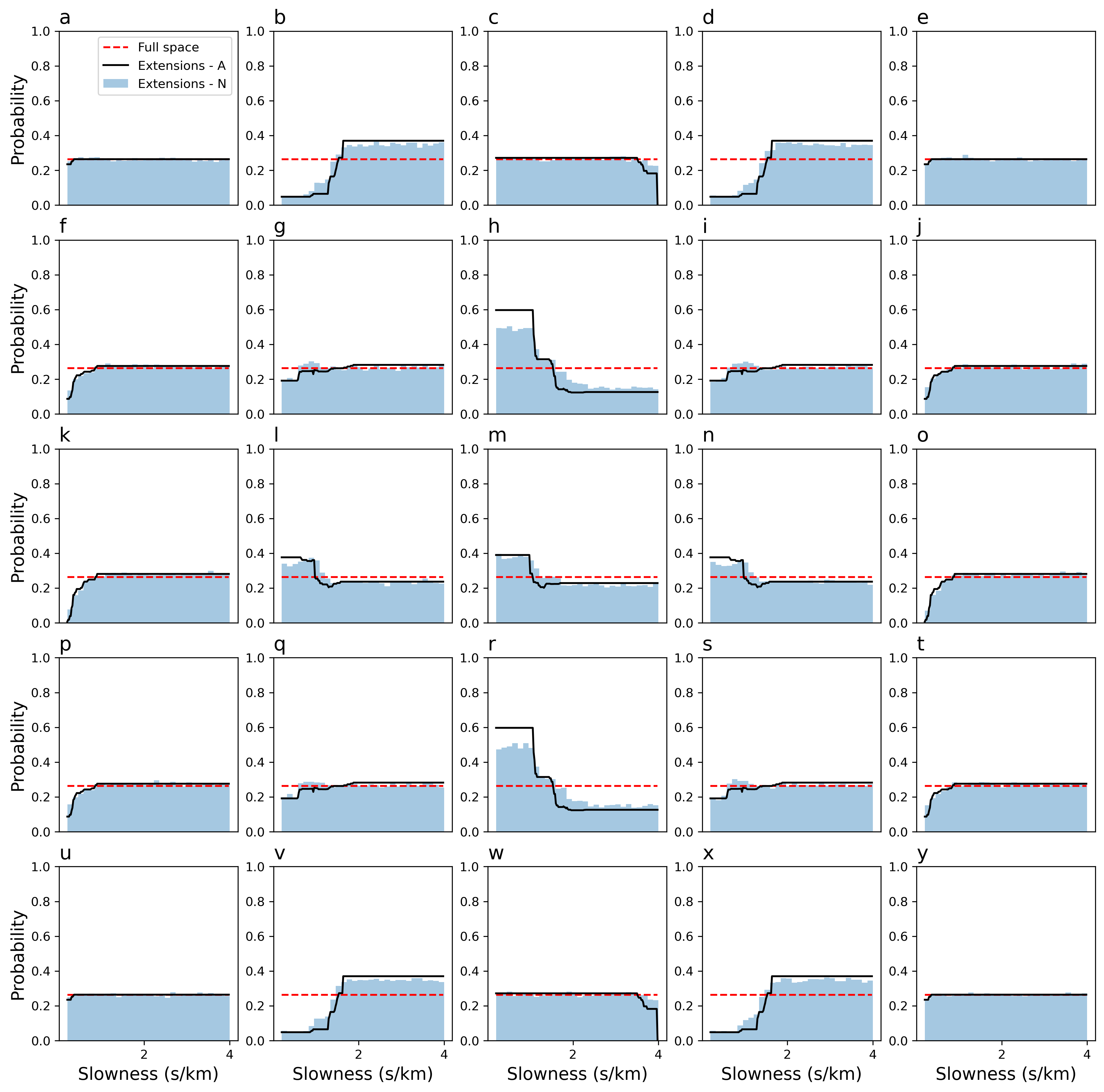}
	\caption{Prior marginal pdf's of the slowness values in the 25 cells. Blue histograms show the prior marginal pdf's obtained by random sampling inside the extensions, black lines show the marginal distributions calculated analytically using equation \ref{eq:prior_allext}, and dashed red lines show the non-zero section of the uniform prior distribution in the full parameter space.}
	\label{fig:prior_full_ext}
\end{figure}

The probability value of a uniform prior distribution is the same everywhere inside the extensions. If we only consider prior parameter space inside one specific extension $\Omega_{k}$ (where subscript $k$ indexes different extensions), the uniform prior distribution can be written as
\begin{equation}
	p_{k}(\mathbf{m}) = 
	\begin{cases}
		\dfrac{1}{V_k} & \mathbf{m} \in \Omega_{k} \\
		0 & \mathbf{m} \notin \Omega_{k}
	\end{cases}
	\label{eq:prior_ext}
\end{equation}
where $V_k$ is the hypervolume of the extension subspace $\Omega_k$ and can be calculated analytically since we know its upper and lower slowness bounds. The total extensions subspace $\Omega_{ext}$ provided by all 709 extensions can be expressed as
\begin{equation}
	\Omega_{ext} = \Omega_1 \cup \Omega_2 \cup ... \Omega_{708} \cup \Omega_{709}
	\label{eq:extensions_union}
\end{equation}
where symbol $\cup$ stands for the union of sets. If we ignore overlapping regions between different extensions, the prior distribution over only the hypervolume spanned by these extensions can be written
\begin{equation}
	p(\mathbf{m}) = 
	\begin{cases}
		\dfrac{\sum_k V_k \ p_{k}(\mathbf{m})}{\sum_k V_k} & \mathbf{m} \in \Omega_{ext} \\
		0 & \mathbf{m} \notin \Omega_{ext} 
	\end{cases}
	\label{eq:prior_allext}
\end{equation}
which can be calculated analytically. Note that any overlapping regions might be repeated multiple times when applying equation \ref{eq:prior_allext}, and this potentially introduces errors into the analytically calculated prior distribution.

Given the relatively low dimensionality (25) of this problem it is feasible to sample inside the extensions to estimate the extent to which the prior distribution is now known (spanned) by the hypervolume of these extensions. This approach would be important in practical cases if the prior distribution was not explicit, so had to be evaluated numerically for each sample; in this case however, it simply provides an opportunity to validate the analytic approximation to the prior pdf given by equation \ref{eq:prior_allext}. In Figure \ref{fig:prior_full_ext}, we compare prior marginal distributions of the slowness values in the 25 grid cells. In each figure, a black line displays the analytically calculated marginal prior pdf (equation \ref{eq:prior_allext}), a blue histogram shows the results of numerical sampling inside the extensions, and a dashed red line shows the true uniform prior distribution in the full parameter space. Comparing the blue histograms to the black lines, the analytic and numerical sampling results are shown to be very similar for most grid cells. The most extreme deviations occur for cells in Figures \ref{fig:prior_full_ext}h and \ref{fig:prior_full_ext}r which display some discrepancies, yet these are still potentially acceptable levels depending on the application. Assuming that the Monte Carlo sampling was sufficiently dense that we can assume the blue histograms to be correct, the only explanation for deviations is due to any overlaps between hypervolumes of different extensions, since these were ignored in equation \ref{eq:prior_allext}. Figure \ref{fig:prior_full_ext} therefore shows that overlaps between different extensions introduce little error, and will therefore be ignored in what follows.

\subsection{Semi-analytic Calculation of Posterior PDF using Extensions}
Bayesian seismic tomographic problems are solved within a probabilistic framework using Bayes' theorem:
\begin{equation}
	p(\mathbf{m}|\mathbf{d}) = \dfrac{p(\mathbf{d}|\mathbf{m})p(\mathbf{m})}{p(\mathbf{d})}
	\label{eq:bayes}
\end{equation}
Here $p(\mathbf{m})$ is the \textit{prior} probability density function (pdf) of model parameter $\mathbf{m}$ and describes the prior knowledge about $\mathbf{m}$ that existed before data $\mathbf{d}$ were introduced. The conditional probability $p(\mathbf{d}|\mathbf{m})$ is called the \textit{likelihood} and is the probability of observing data $\mathbf{d}$ given a particular model $\mathbf{m}$ (that is, under the assumption that model $\mathbf{m}$ is true). Term $p(\mathbf{d})$ is a normalisation constant called the \textit{evidence} which ensures that the expression on the right integrates to 1 and hence produces a valid probability distribution. On the left hand side, $p(\mathbf{m}|\mathbf{d})$ is the so-called \textit{posterior} pdf, that is the probability of model parameter $\mathbf{m}$ being true given the observed data $\mathbf{d}$.

Since the fastest ray path remains unchanged for all model samples within an extension, the forward function which calculates the first arrival travel time through a given slowness model can be written as in equation \ref{eq:tt_cal1}. Clearly, the forward function value only depends on the on-ray slowness values (corresponding to cells whose ray path lengths $l^j > 0$). Off-ray slowness values (corresponding to cells with $l^k = 0$) are identified as redundant parameters for this forward function. Within each extension, travel time prediction thus becomes a linear problem. Note that this differs fundamentally from linearisation of the forward function in a typical locally-linearised inversion, even though equation \ref{eq:tt_cal1} may be the same in both cases: when using extensions, the relationship is obtained from symmetries and other properties of the physical system and is valid over a known and potentially large hypervolume in parameter space defined in known directions away from each sample. On the other hand, when using a first-order expansion to approximate a problem by linearisation, the same equation is assumed to be true only locally, and in all parameter-space directions around each sample. 

Given the approximate prior information obtained previously, we calculate the posterior pdf inside the set of extensions. Consider the tomography problem shown in Figure \ref{fig:ray_paths}, where only 1 source-receiver pair is used so that there is only one travel time datum. Assume a Gaussian likelihood function around the observed datum $d$
\begin{equation}
	p(d|\mathbf{m}) = c \ \exp \left(-\dfrac{(d-\mathbf{l}^T\mathbf{m})^2}{2\sigma ^2}\right)
	\label{eq:likelihood_ext}
\end{equation}
where $c$ is a normalisation constant and $\sigma$ is the estimated uncertainty of this datum. For fixed observed datum $d$, equation \ref{eq:likelihood_ext} defines a non-normalised probability distribution over $\mathbf{m}$, which is the likelihood function. Substituting equations \ref{eq:prior_ext} and \ref{eq:likelihood_ext} into equation \ref{eq:bayes} we obtain
\begin{equation}
	p_k(\mathbf{m}|d) = 
	\begin{cases}
		\dfrac{\exp \left(-\frac{(d-\mathbf{l}_k^T\mathbf{m})^2}{2\sigma ^2}\right)}{\int_{\mathbf{m}\in \Omega_k} \exp \left(-\frac{(d-\mathbf{l}_k^T\mathbf{m})^2}{2\sigma ^2}\right) d\mathbf{m}} & \mathbf{m} \in \Omega_{k} \\
		0 & \mathbf{m} \notin \Omega_{k} 
	\end{cases}
	\label{eq:posterior_ext}
\end{equation}
where $\mathbf{l}_k$ is the ray path vector corresponding to a specific extension subspace $\Omega_k$. Since the forward function is linear within each extension (equation \ref{eq:tt_cal1}), the posterior distribution in equation \ref{eq:posterior_ext} can be calculated analytically.

\begin{figure}
	\centering\includegraphics[width=\textwidth]{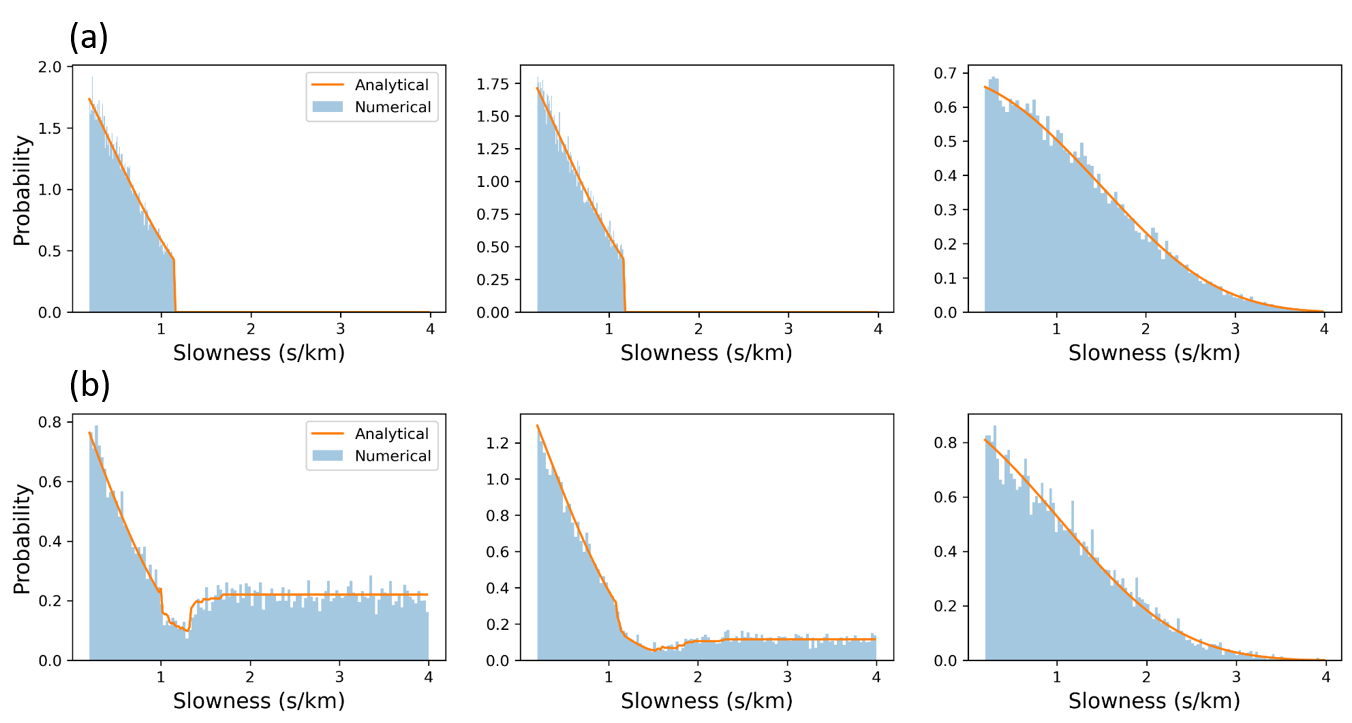}
	\caption{Posterior marginal distributions obtained from (a) a single extension subspace, and (b) all 709 extensions provided by the 51 optimal slowness model samples. In each figure, the orange line is the marginal distribution obtained by analytic calculation and the blue histogram represents equivalent results obtained numerically using McMC.}
	\label{fig:ana_num_posterior_ext}
\end{figure}

We compute the analytic posterior pdf inside a single extension represented by the optimal sample and the straight ray path shown in Figure \ref{fig:optimal_sample_1}. The observed source-to-receiver travel time is $d = 2.9$ s with a corresponding observational uncertainty of $\sigma = 0.1$ s. Given this ray configuration, only prior information about the 5 on-ray cells will be updated by the observed datum. Figure \ref{fig:ana_num_posterior_ext}a displays 3 posterior marginal pdf's for cells in the blue box in Figure \ref{fig:optimal_sample_1} since the top two cells traversed by the ray exhibit identical marginal pdf's to those of the bottom two cells due to the geometrical symmetry inherent in this problem. Orange lines show the posterior marginal distributions obtained from the analytic calculation and blue histograms show the numerical results obtained by running a Markov chain Monte Carlo (McMC) inversion inside this extension. Given that the analytic and numerical approaches yield the same marginal pdf's, we are confident that the analytic calculation is correct.

We also substitute equations \ref{eq:prior_allext} and \ref{eq:likelihood_ext} into equation \ref{eq:bayes} to calculate the posterior pdf inside all extensions. Figure \ref{fig:ana_num_posterior_ext}b shows the obtained posterior marginal pdf's for the 3 grid cells within the blue box in Figure \ref{fig:optimal_sample_1}. Again, orange lines show the analytic results and blue histograms show the McMC results. Once more, the analytic method provides accurate posterior marginal pdf's that agree with the McMC results.

\begin{figure}
	\centering\includegraphics[width=\textwidth]{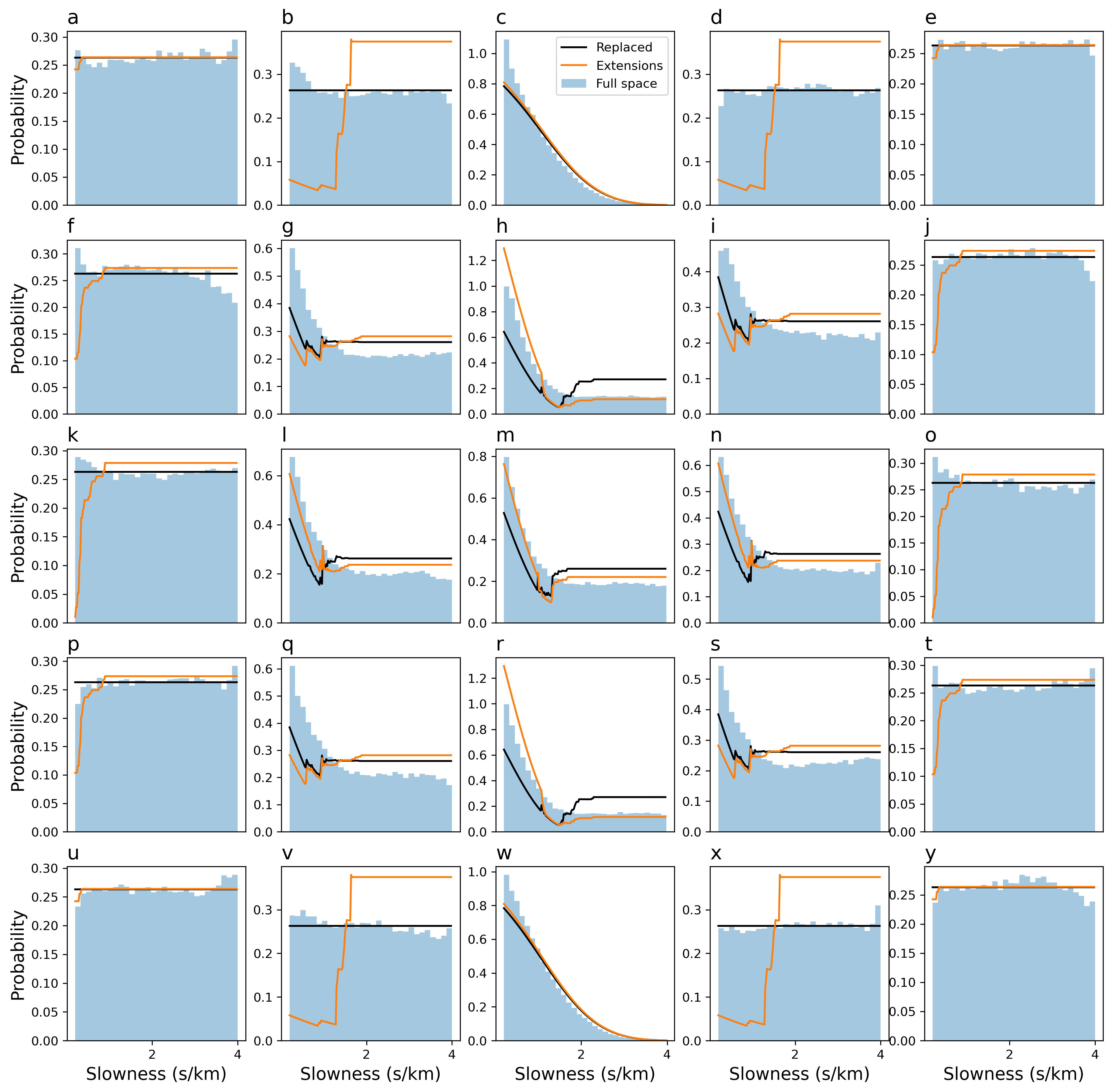}
	\caption{Posterior marginal distributions of the slowness values in all 25 cells of the grid in Figure \ref{fig:ray_paths}. Orange lines show the marginal pdf's obtained using analytic calculations inside the extensions, black lines show those obtained after applying prior replacement, and blue histograms show those from McMC inversion performed in the full parameter space.}
	\label{fig:ext_full_posterior}
\end{figure}

Figure \ref{fig:ext_full_posterior} displays the posterior marginal pdf's for all 25 grid cells. Orange lines show marginal pdf's calculated analytically. We also perform a McMC inversion in the full parameter space. Note that this is different from previous McMC tests displayed in Figure \ref{fig:ana_num_posterior_ext}b which are performed exclusively within extensions subspaces. The corresponding marginal pdf's are represented by blue histograms in Figure \ref{fig:ext_full_posterior}. This McMC result is taken as the reference solution for this single-datum tomography inversion, for comparison with the analytic results calculated using extensions. For some cells, the analytic results obtained using extensions align with the sampling results performed in the full parameter space. However, marginal pdf's displayed in Figures \ref{fig:ext_full_posterior}b, d, g, i, q, s, v and x exhibit significant discrepancies. 

One contribution to the errors in Figure \ref{fig:ext_full_posterior} (between the orange lines and blue histograms) is the presence of bias in the prior information provided by the extensions (black lines in Figure \ref{fig:prior_full_ext}). This bias is particularly prominent for cells shown in Figures \ref{fig:prior_full_ext}b, d, h, r, v and x, in comparison to the uniform prior distribution across the entire parameter space (dashed red lines in Figure \ref{fig:prior_full_ext}). To address this bias, we apply a \textit{prior replacement} method \cite{walker2014varying, zhao2024variational}. This technique enables prior information to be updated either during or after Bayesian inversion. In accordance with Bayes' theorem, the posterior distribution following prior replacement can be represented as:
\begin{equation}
	\begin{split}
		p_{new}(\mathbf{m}|\mathbf{d}) & = \dfrac{p(\mathbf{d}|\mathbf{m})p_{old}(\mathbf{m})}{p_{old}(\mathbf{d})} \  \dfrac{p_{new}(\mathbf{m})}{p_{old}(\mathbf{m})} \ \dfrac{p_{old}(\mathbf{d})}{p_{new}(\mathbf{d})} \\
		& = b \ p_{old}(\mathbf{m}|\mathbf{d}) \ \dfrac{p_{new}(\mathbf{m})}{p_{old}(\mathbf{m})}
	\end{split}
	\label{eq:prior_replacement}
\end{equation}
where $b=p_{old}(\mathbf{d})/p_{new}(\mathbf{d})$ is a normalisation constant. Subscripts \textit{old} and \textit{new} stand for terms before and after updating the prior pdf. By dividing the biased posterior distribution $p_{old}(\mathbf{m}|\mathbf{d})$ by the biased prior distribution $p_{old}(\mathbf{m})$ and multiplying by the unbiased prior pdf $p_{new}(\mathbf{m})$ (which corresponds to the uniform prior distribution in the full parameter space), we can correct the bias introduced by the old prior information.

We implement the prior replacement by dividing the orange lines in Figure \ref{fig:ext_full_posterior} by the black lines in Figure \ref{fig:prior_full_ext} and then multiplying by the dashed red lines in Figure \ref{fig:prior_full_ext}. Note that this approach disregards any potential correlations between different model parameters. The corresponding results are displayed by black lines in Figure \ref{fig:ext_full_posterior}. Compared to the orange lines, the posterior marginal distributions for some cells are improved by prior replacement (e.g., cells displayed in Figures \ref{fig:ext_full_posterior}b, d, v and x). However, for some cells the posterior marginal pdf's following the prior replacement become less accurate (Figures \ref{fig:ext_full_posterior}h and \ref{fig:ext_full_posterior}r). Even though the results are not perfect we note that both results are obtained by analytic calculations, and employ only 51 deterministically chosen samples. In contrast, the McMC results are obtained using 500,000 samples (and thus 500,000 forward evaluations). In the next section, we will use the analytic results together with prior replacement (as denoted by the black lines in Figure \ref{fig:ext_full_posterior}) to compute Bayesian tomographic results that take into account multiple travel time data constraints.

\subsection{Travel Time Tomography with Multiple Data}
In (Bayesian) inverse problems it is often the case that multiple data points are assumed to be mutually independent, and to have a symmetric uncertainty around the observed value with a spread that depends on the distance between observed and modelled data. A Gaussian distribution with diagonal covariance matrix is often used as the likelihood function:
\begin{equation}
	\begin{split}
		p(\mathbf{d}|\mathbf{m}) &= c \ \exp \left(-\sum_i \dfrac{(d_i - \mathbf{l}_i^T\mathbf{m})^2} {2\sigma_i^2}\right) = \prod_i c_i \ \exp \left(-\dfrac{(d_i - \mathbf{l}_i^T\mathbf{m})^2} {2\sigma_i^2}\right) \\
		& = \prod_i p(d_i|\mathbf{m})
	\end{split}
	\label{eq:likelihood}
\end{equation}
where $\sigma_i$ is the uncertainty for the $i$th datum $d_i$. The second line of equation \ref{eq:likelihood} shows that the full likelihood function can be represented as the product of individual likelihoods $p(d_i|\mathbf{m})$, each of which corresponds to a sub-problem involving a different single datum $d_i$ and the same model parameter $\mathbf{m}$ \cite{curtis2020samples}. Given a uniform prior distribution, \citet{curtis2020samples} reformulated Bayes' rule (equation \ref{eq:bayes}) as
\begin{equation}
	p(\mathbf{m}|\mathbf{d}) = \dfrac{p(\mathbf{m})}{p(\mathbf{d})} \prod_i p(d_i|\mathbf{m}) 
	= k' \ \prod_i p(\mathbf{m}|d_i)
	\label{eq:bayes_sub}
\end{equation}
where the second equality is obtained by Bayes' rule applied to likelihood $p(d_i|\mathbf{m})$, and $k'$ is a normalisation constant to ensure that the result on the right side is a valid probability distribution. In this way the solution of a full Bayesian problem can be constructed by combining the posterior distributions of multiple Bayesian sub-problems, each involving constraints from a single datum. Using the mean field approximation (which states that posterior model parameter uncertainties are independent of one another) which is commonly applied in many fields \cite{bishop2006pattern, kucukelbir2017automatic, nawaz2018variational}, equation \ref{eq:bayes_sub} can be approximated by
\begin{equation}
	p(m_t|\mathbf{d}) \approx k_t' \ \prod_i p(m_t|d_i)
	\label{eq:bayes_sub_marginal}
\end{equation}
where $m_t$ is an arbitrary element in model parameter vector $\mathbf{m}$. Equation \ref{eq:bayes_sub_marginal} states that the posterior marginal pdf's for the full inversion results constrained by multiple data can be constructed approximately by a collection of marginal pdf's, each derived from an individual travel time datum. In the previous section, these individual marginal pdf's have been calculated analytically. Hence, equation \ref{eq:bayes_sub_marginal} can also be calculated analytically.

\begin{figure}
	\centering\includegraphics[width=0.5\textwidth]{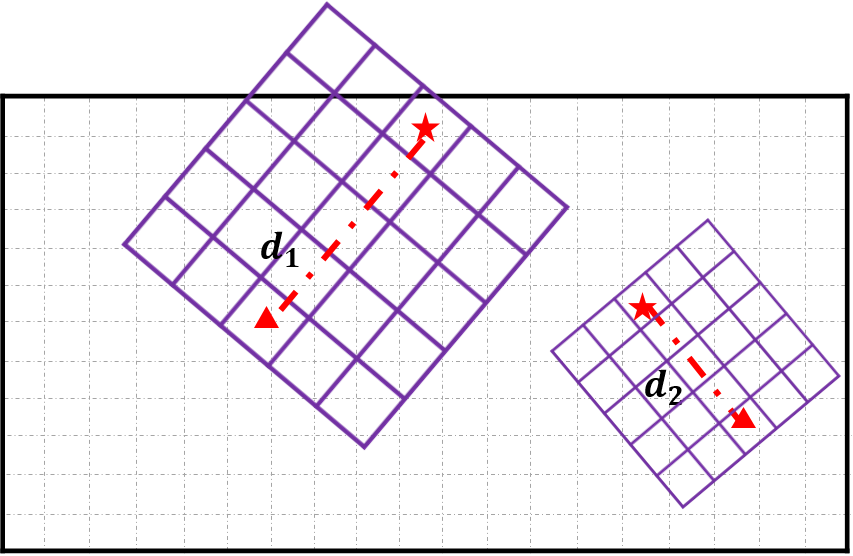}
	\caption{Schematic diagram illustrating the construction of the posterior marginal pdf's of the full inversion results (parametrised by background grey grid cells) given multiple travel times, based on the marginal pdf's of the inversion results obtained from each individual travel time datum. Red stars and triangles show the locations of 2 sources and 2 receivers, between which we obtain 2 source-to-receiver travel times. We use the same $5\times 5$ regular grid system as displayed in Figure \ref{fig:ray_paths} to parametrise the imaging region around each source-receiver pair. For example, in the case of the source-receiver pairs denoted by $d_1$ and $d_2$ as depicted here, 2 purple grid systems show the parametrised grid systems. For spatial locations inside the parametrised $5 \times 5$ grids, their posterior marginal pdf's are updated using equation \ref{eq:update_marginal_inside}. Otherwise we keep them unchanged (equation \ref{eq:update_marginal_outside}).}
	\label{fig:multiple_data_illustration}
\end{figure}

Figure \ref{fig:multiple_data_illustration} illustrates a procedure used for the above analytic calculation. The imaging region (black rectangle) is parametrised into a set of regular grid cells (dashed grey ones). Two sources (red stars) and two receivers (red triangles) are placed within the imaging region and provide two source-receiver travel times $d_1$ and $d_2$. We wish to calculate the posterior marginal pdf's of slowness values of the grey cells given the observed travel time dataset. For each source-receiver pair, its surrounding spatial (imaging) domain is parametrised using the same $5\times 5$ regular grid system as that in Figure \ref{fig:ray_paths} (with necessary scale and rotation of the grid system). For example, 2 purple grid systems in Figure \ref{fig:multiple_data_illustration} display the parametrised regions established around the source-receiver pairs represented by $d_1$ and $d_2$. Based on the source-to-receiver distance of each specific pair, we scale the ray path matrix $L$. Notably from equation \ref{eq:optimiser_ipm}, the optimisation results (the optimal samples) for different source-receiver pairs are exactly the same if the lower and upper bounds of the prior distributions are the same for all spatial locations (which is common in seismic travel time tomography problems) and if the (scaled) ray path matrices for various source-receiver pairs are defined using the same template grid system (in this work, the same $5\times 5$ grid system displayed in Figure \ref{fig:ray_paths} is used). Therefore, the optimisation problem in equation \ref{eq:optimiser_ipm}, as well as the single-datum analytic posterior pdf, needs to be solved only once. These results can then be scaled, rotated and used for different travel times.

We substitute each travel time datum, the scaled ray matrix $L$ and the optimal samples into equation \ref{eq:posterior_ext}. This allows us to calculate the inversion results for a single datum analytically. To mitigate the impact of the biased prior information from the extensions, we apply the prior replacement method outlined in equation \ref{eq:prior_replacement}. These single-datum posterior marginal pdf's are used to create full inversion results constrained by multiple data, using equation \ref{eq:bayes_sub_marginal}. To explain, from Figure \ref{fig:multiple_data_illustration} each travel time datum can only update information of spatial locations where the imaging area (dashed grey cells) coincides with the $5\times 5$ grid system (purple grids). For any grey grid cell $m_t$ if its central location is inside the $5\times 5$ purple grid system defined by datum $d_i$, we update its posterior marginal pdf using
\begin{equation}
	p(m_t|\mathbf{d}_{1:i}) = k_t' \ p(m_t|\mathbf{d}_{1:i-1}) \ p(m_t|d_{i})
	\label{eq:update_marginal_inside}
\end{equation}
where $\mathbf{d}_{1:i}$ (or $\mathbf{d}_{1:i-1}$) stands for the first $i$ (or $i-1$) travel time data. Otherwise if the central location of cell $m_t$ is outside the $5\times 5$ grid system, we keep its posterior marginal pdf unchanged
\begin{equation}
	p(m_t|\mathbf{d}_{1:i}) = p(m_t|\mathbf{d}_{1:i-1})
	\label{eq:update_marginal_outside}
\end{equation}
The latter equation implies that the considered datum $d_i$ makes no contribution to the imaging result of the spatial location represented by $m_t$. We repeat this procedure for each source-receiver pair (each travel time datum $d_i$), and obtain the posterior marginal pdf's of the full inversion results. This gives a new extensions-based tomography algorithm as follows:
\begin{enumerate}
	\item Parametrise the local region around a given source-receiver pair into a $5 \times 5$ regular grid system.
	\item Scale the ray matrix based on the source-to-receiver distance.
	\item Calculate the analytic posterior marginal pdf's for the considered source-receiver pair using equations \ref{eq:posterior_ext} and \ref{eq:prior_replacement}.
	\item Update the posterior marginal pdf's of the full inversion results at every spatial location using equation \ref{eq:update_marginal_inside} or \ref{eq:update_marginal_outside}.
	\item Repeat (i) -- (iv) for all source-receive pairs.
\end{enumerate}
Based on this algorithm, we are able to calculate any first-order statistical information of the posterior distribution analytically for a travel time tomography problem using sample extensions.

\section{Examples}
\subsection{Toy example}
\begin{figure}
	\centering\includegraphics[width=0.4\textwidth]{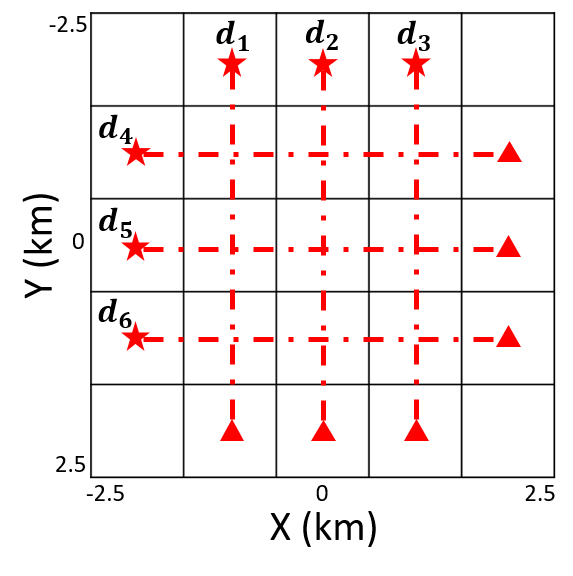}
	\caption{Geometrical configuration used for the conceptual tomography example, which contains 6 observed travel time data. Sources and receivers are represented by red stars and triangles, respectively. Dashed red lines indicate 6 travel time data considered in this example.}
	\label{fig:geometry_6data}
\end{figure}

In this section, we apply the proposed extensions-based algorithm to two travel time tomography examples. In the first example, we test the above method by solving a conceptual tomography problem that contains 6 travel time data, as shown in Figure \ref{fig:geometry_6data}. Red stars stand for the locations of 6 sources and red triangles stand for those of 6 receivers. Dashed red lines that connect source-receiver pairs represent the 6 travel times considered in this example. We focus on these particular travel times to maintain the simplicity of this first problem.

\begin{figure}
	\centering\includegraphics[width=\textwidth]{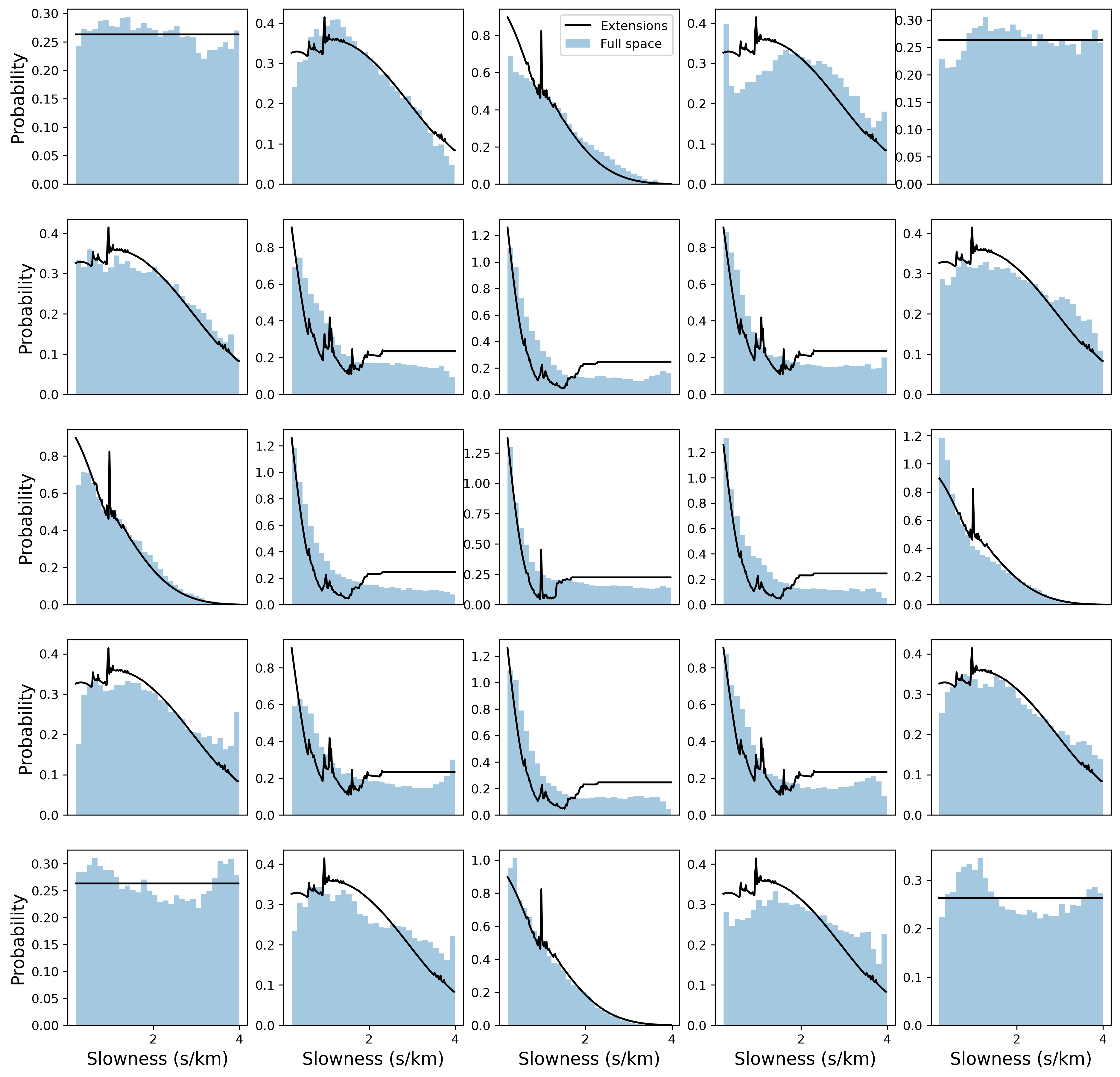}
	\caption{Posterior marginal distributions of the slowness values for the 25 grid cells displayed in Figure \ref{fig:geometry_6data}. Black lines show the marginal pdf's obtained using the proposed extensions-based algorithm, and blue histograms show those from McMC sampling performed in the full parameter space.}
	\label{fig:ext_full_posterior_6data}
\end{figure}

In this example, we set the 6 travel time data to have equal values of 2.9 s and use them to estimate the posterior pdf of the slowness field within the region shown in Figure \ref{fig:geometry_6data}. We do not define a \textit{true} slowness model explicitly and only compare the results obtained using the proposed extensions-based method and Monte Carlo sampling, given the 6 data. We parametrise the slowness field to be estimated using the same $5\times 5$ regularly gridded system as we used in Figure \ref{fig:ray_paths}. This allows us to treat the analytic posterior marginal pdf's obtained in the previous section (black lines in Figure \ref{fig:ext_full_posterior}) as the results for $d_2$. We calculate the analytic posterior marginal pdf's for all of the 6 data points, and combine them to get the results shown in Figure \ref{fig:ext_full_posterior_6data}. Black lines are the resulting analytic marginal pdf's of slowness values in the 25 grid cells. We also perform a McMC inversion in the full parameter space by drawing 1 million samples, and the corresponding marginal pdf's are displayed by the blue histograms in Figure \ref{fig:ext_full_posterior_6data}. 

In most instances, the analytic marginal pdf's obtained using the extensions-based algorithm are similar to those obtained from McMC, which serves as the reference solution for this tomography problem. It is worth noting that while some inconsistencies can be observed for specific cells, these discrepancies may be at an acceptable level, given that the analytic results are obtained using only 51 deterministic samples and their extensions while the Monte Carlo (numerical) results are obtained using 1 million samples. This example demonstrates that the proposed algorithm is able to provide approximately correct first-order statistics of the inversion results. However, the method fails to capture higher-order statistics, because to simplify analytic expressions we applied the mean field approximation which ignores correlations between different model parameters.

\subsection{Synthetic Tomography Example}
\begin{figure}
	\centering\includegraphics[width=0.4\textwidth]{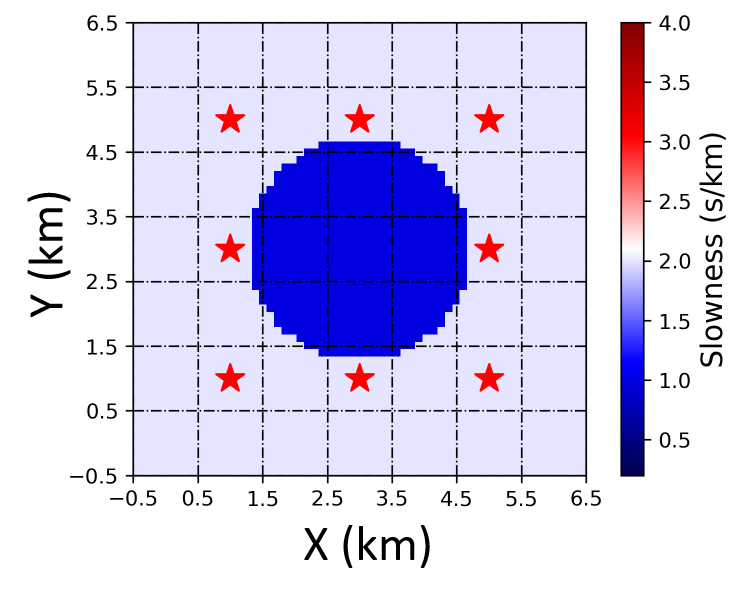}
	\caption{True slowness model for the 2D synthetic tomography test. A low slowness circular anomaly with slowness 1 $s/km$ is embedded within a background slowness of 2 $s/km$. Red stars show the locations of 8 receivers (and sources), and travel times between each pair of locations form the observed data set (28 travel times) in this example.}
	\label{fig:tomo_low_slow_true}
\end{figure}

In this second example, we consider a slightly more complex 2D travel time tomography scenario. Figure \ref{fig:tomo_low_slow_true} shows the true slowness model, which consists of a circular low slowness anomaly of 1 s/km surrounded by a high slowness background region of 2 s/km. Red stars show the locations of 8 co-located sources and receivers placed around the central anomaly, and between each source-receiver pair at distinct locations we obtain a first arrival travel time. We thus obtain 28 independent travel times, which form the observed data set for this problem. For inversion we parametrise the slowness vector $\mathbf{m}$ into 7 $\times$ 7 regular grid cells with a grid size of 1 km in both directions. We define a uniform prior probability distribution bounded between 0.2 and 4.0 s/km for each grid cell. The likelihood function is chosen to be a diagonal Gaussian distribution with a data uncertainty of $\sigma_d = 0.1$ s for all data points.

\begin{figure}
	\centering\includegraphics[width=\textwidth]{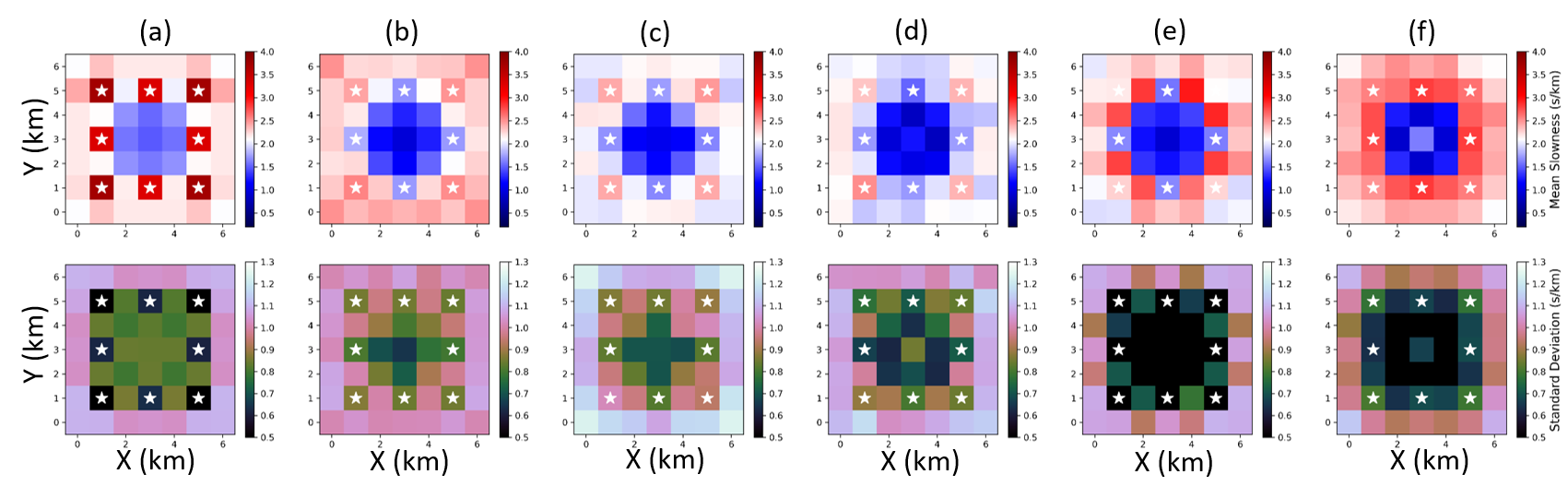}
	\caption{Inversion results for the 2D synthetic tomography example using different methods. Top row shows the mean slowness model and bottom row shows the standard deviation model of the posterior distribution. White stars display source and receiver locations. Inversion results obtained using (a) the proposed extensions-based tomography algorithm and (f) MH-McMC with FMM employed to solve the forward function. Details about other panels can be found in the main text.}
	\label{fig:tomo_low_slow_mean_std}
\end{figure}

\begin{figure}
	\centering\includegraphics[width=\textwidth]{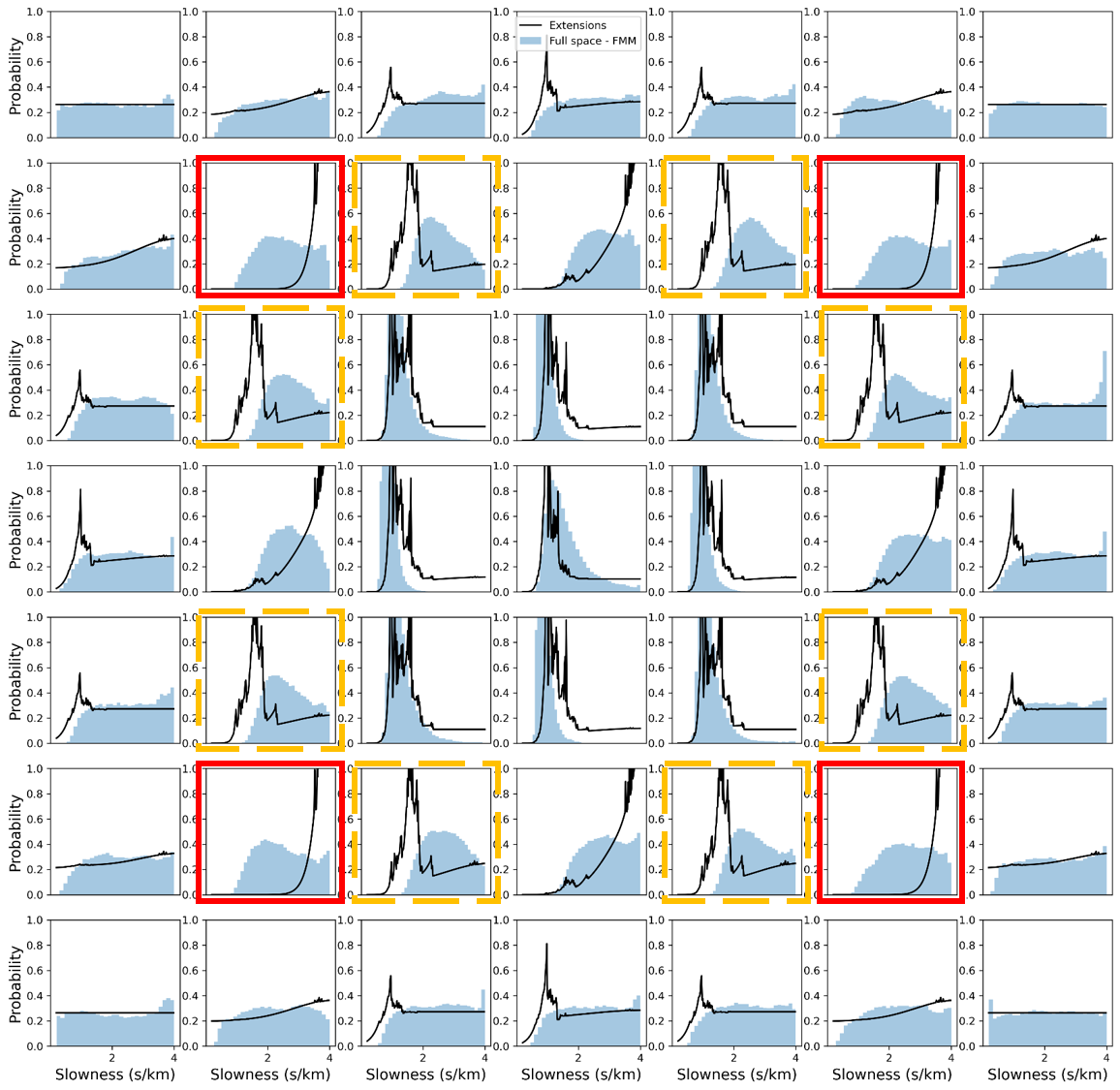}
	\caption{Posterior marginal pdf's of the slowness values in each of the 7 $\times$ 7 grid cells. In each figure, the black line shows the marginal distribution obtained using the extensions-based algorithm (corresponding to the inversion results displayed in Figure \ref{fig:tomo_low_slow_mean_std}a), and the blue histogram shows that obtained from MH-McMC (corresponding to Figure \ref{fig:tomo_low_slow_mean_std}f). Red and dashed yellow boxes indicate cells where the extensions-based algorithm yields notably biased results compared to MH-McMC.}
	\label{fig:tomo_low_slow_marginal}
\end{figure}

Within this imaging region, we first use the proposed extensions-based algorithm to solve this tomography problem. As depicted in Figure \ref{fig:tomo_low_slow_marginal}, the resultant posterior marginal pdf's are represented by the black lines. Based on these marginal pdf's, we calculate various first-order statistical properties of the posterior distribution analytically. For example, Figure \ref{fig:tomo_low_slow_mean_std}a shows the mean (top row) and standard deviation (bottom row) maps calculated from the marginal pdf's in Figure \ref{fig:tomo_low_slow_marginal}. The mean model identifies and reconstructs the low slowness anomaly inside the receiver array effectively. Due to the relatively coarse grid of cells employed in this example, the spatial resolution remains limited and the result fails to capture the circular shape of the anomaly precisely. Notably at the station locations, the mean slownesses exhibit higher values, while their corresponding uncertainties are relatively lower compared to the surrounding cells.

To validate the above results, we run a Metropolis-Hastings Markov chain Monte Carlo (MH-McMC) test for comparison. We employ the widely adopted fast marching method \cite[FMM -- ][]{rawlinson2005fast} to solve the forward problem and predict synthetic travel times. We run 4 Markov chains in parallel, each drawing 250,000 samples. After sampling, we discard the first 100,000 samples as the burn-in period, and retain every 30th sample from the remaining samples to approximate the slowness samples of the posterior distribution. The sampling result serves as a reference solution for this tomographic problem. The mean and standard deviation maps are displayed in Figure \ref{fig:tomo_low_slow_mean_std}f, and the corresponding marginal pdf's are displayed by the blue histograms in Figure \ref{fig:tomo_low_slow_marginal}. We observe that the mean models in Figures \ref{fig:tomo_low_slow_mean_std}a and \ref{fig:tomo_low_slow_mean_std}f show some similar features. For example, both methods recover the central anomaly, albeit with a relatively low resolution. The surrounding region exhibits higher slowness values in both cases. The two methods also provide similar posterior marginal pdf's for the 9 central cells. However, some inconsistencies between these two results are observed in the mean and standard deviation maps, as well as in the marginal pdf's in Figure \ref{fig:tomo_low_slow_marginal}. Four red boxes in Figure \ref{fig:tomo_low_slow_marginal} highlight cells whose posterior marginal distributions show substantial discrepancies between these two results, and dashed yellow boxes highlight cells with moderate discrepancies. Since the results from McMC can be considered as reference results for a nonlinear problem provided that the chains have converged sufficiently, we think that the results from extensions-based algorithm are likely to be the more biased. Nevertheless, these findings still offer valuable insights into the true solution, particularly considering that they are obtained in an analytic manner from 51 deterministic samples, compared to one million random samples used in the Monte Carlo test.

To investigate potential sources of errors that may contribute to the bias observed in Figures \ref{fig:tomo_low_slow_mean_std}a and \ref{fig:tomo_low_slow_marginal}, we make a series of tests to analyse approximations employed in the extensions-based algorithm. First, given that the algorithm operates within the extensions subspace rather than the full parameter space as depicted in Figure \ref{fig:extension_illustration}b (the blue subspace compared to the full space), this must introduce bias to the final results. To test this, we conduct a Monte Carlo test within the full parameter space for each travel time datum, and use the results to calculate single-datum posterior marginal pdf's numerically. We use the (numerically) obtained single-datum marginal pdf's to construct a full inversion result encompassing all 28 travel times. This is achieved using exactly the same procedure as that used in the extensions-based algorithm, which includes applying prior replacement, updating full posterior marginal pdf's, and iterating for all data points. This ensures that the results are directly comparable to those in Figure \ref{fig:tomo_low_slow_mean_std}a, the only difference being that the Monte Carlo results consider the whole parameter space rather than the extensions subspace. The mean and standard deviation maps from this Monte Carlo test are displayed in Figure \ref{fig:tomo_low_slow_mean_std}b. Since we observe significant differences outside of the low slowness anomaly in both mean and standard deviation maps in Figures \ref{fig:tomo_low_slow_mean_std}a and \ref{fig:tomo_low_slow_mean_std}b, we deduce that the limited extensions subspace indeed introduces notable errors into the inversion results.

Significant differences still exist between Figures \ref{fig:tomo_low_slow_mean_std}b and \ref{fig:tomo_low_slow_mean_std}f, which indicates the presence of additional sources of error that contribute to the biased results in Figures \ref{fig:tomo_low_slow_mean_std}a and \ref{fig:tomo_low_slow_mean_std}b. In both of those results we used 61 rays (displayed in Figure \ref{fig:ray_paths}) to obtain the optimal samples (and their extensions) and to calculate the forward function values via equation \ref{eq:tt_cal_matrix}. To ascertain whether this set of sparsely defined rays sufficiently captures the contributions from all possible ray paths between source and receiver locations, we define a new ray matrix that contains 5000 different rays between these two locations. These rays are selected by drawing random slowness samples from the prior distribution and calculating the corresponding fastest rays between the source and receiver locations displayed in Figure \ref{fig:ray_paths}. We perform another analogous test to that presented in Figure \ref{fig:tomo_low_slow_mean_std}b, with the sole change being the use of this new ray matrix to perform Monte Carlo (numerical) inversion (note that in these two tests we perform numerical inversion rather than analytic calculations). The result is displayed in Figure \ref{fig:tomo_low_slow_mean_std}c. We obtain very similar mean and standard deviation maps inside the station array, which suggests (perhaps surprisingly) that the limited number of rays (61) introduces only minor errors to the tomographic solution.

In the extensions-based algorithm, we parametrise the spatial region around each source-receiver pair into the same $5 \times 5$ regular gridded system as depicted in Figure \ref{fig:ray_paths}, such that the deterministic samples obtained by solving the optimisation problem in equation \ref{eq:optimiser_ipm} remain the same for all source-receiver pairs. This, however, mandates the need to rotate, shift, and scale the original $5 \times 5$ grid system to align with different source-receiver pairs. To justify the influence of this parametrisation scheme on the final inversion results, in the next test we employ a $7 \times 7$ regular grid discretization for all 28 source-receiver pairs, akin to the discretization used to present the inversion results in Figure \ref{fig:tomo_low_slow_mean_std}. We perform single-datum inversions for each source-receiver pair with the fixed model parametrisation, and use the results to update full inversion using equation \ref{eq:bayes_sub_marginal}, similarly to our preceding tests. Note that in this test the ray matrix also contains 5000 rays to remove potential bias caused by a limited number of rays. The results are displayed in Figure \ref{fig:tomo_low_slow_mean_std}d, and to be clear, the only difference to the method used to produce panel c is the change from $5 \times 5$ to $7 \times 7$ grid cells. By comparing Figures \ref{fig:tomo_low_slow_mean_std}c and \ref{fig:tomo_low_slow_mean_std}d, it is evident that limited number of cells used in parametrisations within the imaging region do influence the final results, but perhaps not quite to the same extent as was observed moving from panels a to b.

To obtain the mean and standard deviation maps in Figures \ref{fig:tomo_low_slow_mean_std}a -- \ref{fig:tomo_low_slow_mean_std}d, we construct the full inversion results obtained from each individual datum using equation \ref{eq:bayes_sub_marginal} as opposed to equation \ref{eq:bayes_sub}. This approach neglects potential correlations among different model parameters, which possibly introduces error. To test this aspect, we run another test that uses all available source-receiver travel times within a single inversion, similarly to conventional Monte Carlo tomography methods. Equation \ref{eq:tt_cal_matrix} is used to solve the forward function, where the ray matrix built by the same 5000 rays is employed. The inversion result is displayed in Figure \ref{fig:tomo_low_slow_mean_std}e. According to equation \ref{eq:bayes_sub}, this result can be interpreted as the combination of a set of full posterior distributions (rather than posterior marginal distributions), each representing an inversion result corresponding to an individual datum. Therefore, the only difference between Figures \ref{fig:tomo_low_slow_mean_std}d and \ref{fig:tomo_low_slow_mean_std}e lies in the application of either equation \ref{eq:bayes_sub} or equation \ref{eq:bayes_sub_marginal} to formulate the full posterior distribution. This discrepancy becomes evident in both the posterior mean and standard deviation maps, which signifies that approximating the posterior pdf by combining only single-parameter marginal pdf's (that is, the mean field approximation) introduces substantial error to the inversion results.

Finally, by comparing the inversion results in Figures \ref{fig:tomo_low_slow_mean_std}e and \ref{fig:tomo_low_slow_mean_std}f, we find that different forward modellers (ray based method in equation \ref{eq:tt_cal_matrix} versus FMM) yield an intermediate level of disparity compared to the above discussed factors. Note that these two methods are based on different physical approximations and both are used in practice \cite{curtis2002probing, rawlinson2005fast, bodin2009seismic, galetti2017transdimensional, fichtner2019hamiltonian}. As such, we refrain from favouring one over the other, as they each have merits and are used widely within the geophysical community.

Table \ref{table:error} presents a summary of the five identified sources of errors outlined in the preceding solution discussion, along with the extent of their influence on the inversion results shown in Figure \ref{fig:tomo_low_slow_mean_std}a. Through systematic examination by altering individual factors in isolation, we have pinpointed two significant factors that impede the efficacy of the proposed approach: (1) limited hypervolume spanned by the extensions; and (2) construction of full inversion results using marginal pdf's obtained from each single-datum inversion. Addressing these two aspects might lead to improvement in the inversion results exhibited in Figure \ref{fig:tomo_low_slow_mean_std}a.

\begin{table}
	\caption{Sources of errors and their effects on the final inversion results.}
	\centering
	\begin{tabular}{cc}
		\hline
		Source of Error & Effect on final results \\
		\hline
		Limited extensions hypervolume  & Large \\
		Limited number of rays  & Negligible \\
		Limited imaging region parametrisation  & Small \\
		Construction of posterior pdf using marginals  & Large \\
		Different forward modellers  & Medium \\
		\hline
	\end{tabular}
	\label{table:error}
\end{table}

\section{Discussion}
We have demonstrated that in principle, sample extensions can provide significant value for travel time tomography. Information obtained from extensions grows exponentially with the number of parameters, almost matching the rate of increasing demand for more sampling due to the curse of dimensionality. This enables efficient exploration of some low-dimensional parameter spaces using only tens of deterministic samples \cite{curtis2020samples}, and results in a new travel time tomography algorithm based on extensions. In our synthetic tomography example, we successfully identified two primary sources of errors that contribute to biased inversion results, as detailed in Table \ref{table:error}. While the extensions subspace covers an average of $87.3\%$ per parameter, the overall hypervolume spanned by extensions in the full parameter space remains limited due to the curse of dimensionality \cite{curtis2001prior}. Meanwhile, the inherent bias in information provided by the off-ray and on-ray extensions is apparent in the sense that extensions are confined to the low slowness side for on-ray cells and the high slowness side for off-ray cells, as depicted in Figure \ref{fig:extension_illustration}b. It is straightforward to recognise that extensions provided by the optimal samples tend to populate hypercorners within the full parameter space (analogous to the upper-left corner of a 2-dimensional space in Figure \ref{fig:extension_illustration}b). The central region of parameter space remains underexplored. This inherent bias could potentially be mitigated by applying methods such as the informed proposal Monte Carlo \cite{khoshkholgh2021informed, khoshkholgh2022full}: by leveraging information derived from extensions we might be able to formulate an informed proposal distribution, which might alleviate the bias through a few Monte Carlo steps, making the results less biased.

The second error arises from the use of equation \ref{eq:bayes_sub_marginal} to combine inversion results obtained from each individual datum. Obviously, marginal pdf's (i.e., the mean field approximation) cannot represent the full posterior distribution accurately, but to apply the more accurate equation \ref{eq:bayes_sub} directly, an analytic expression is required for the full posterior distributions $p(\mathbf{m}|d_i)$. While Gaussian distributions are fully analytic and the product of Gaussians are analytically tractable, the approximation of $p(\mathbf{m}|d_i)$ through a sole Gaussian distribution has been shown to be inaccurate in numerous geophysics studies \cite{zhang2019seismic}. Recently, \citet{zhao2024bayesian} introduced a \textit{boosting variational inference} method to geophysics, which uses a mixture of Gaussian distributions to approximate the true posterior distribution. We might therefore incorporate the principles of boosting variational inference into the proposed extensions-based algorithm: by optimising a mixture of Gaussians to approximate the posterior distribution obtained from each datum, we can combine the full inversion results using equation \ref{eq:bayes_sub}, thus addressing the second type of error.

Extensions are derived from our existing understanding of the underlying physical principles that govern the forward problem. Different forward functions typically correspond to varying physical relationships and, thus lead to distinct extensions. For example, in full waveform inversion the forward problem involves solving the wave equation to predict full waveform data. Consider the 1D scalar wave equation
\begin{equation}
	\dfrac{1}{v(x)^2}\dfrac{\partial^2 p(x,t)}{\partial t^2} = \dfrac{\partial^2 p(x,t)}{\partial x^2}
	\label{eq:wave_equation}
\end{equation} 
where $v(x)$ stands for the acoustic velocity at location $x$. A so-called scale extension exists \cite{curtis2020samples}: if we scale $x$ and $v$ with the same positive number $\lambda$, that is 
\begin{equation}
	\bar x = \lambda \ x, \qquad \bar v = \lambda \ v
	\label{eq:scale_we}
\end{equation}
equation \ref{eq:wave_equation} remains identical, and the modelled full waveform data would be exactly the same. Therefore, we immediately obtain forward function values for many other models, each featuring distinct parametrisations as the spatial domain is scaled by $\lambda$, all from the scale extension of a single model sample \cite{curtis2020samples}. According to the No Free Lunch theorem, it is conceivable that extensions-based algorithms (defined for a limited class of problems with specific physical extensions) might outperform algorithms which are designed to address a broader range of problems (e.g., McMC). One aim of this work is to test this in the case of travel time tomography; further attention needs to be devoted to define and use extensions for other forward (physical) problems. 

\citet{curtis2020samples} introduced the concept of sample extensions and their potential applications in geophysics. Four main types of extensions are defined, namely \textit{function value, data space, parameter space and interpretational extensions}, which might provide useful information for model selection problems \cite{box1987empirical, linde2014falsification}, reversible-jump McMC \cite{green1995reversible, malinverno2002parsimonious}, interrogation problems \cite{poliannikov2016effect, arnold2018interrogation, ely2018assessing, zhao2022interrogating, zhang2021interrogation, siahkoohi2022deep}, experimental design problems \cite{curtis2004theory, curtis2004theory2, guest2009iteratively, bloem2020experimental, strutz2023variational}, as well as geophysical imaging problems as explored both in \citet{curtis2020samples} and the present work. One of our contributions here is to expand the applicability of extensions to slightly more realistic travel time tomography problems than the one considered explicitly in \citet{curtis2020samples}. Moving forward, more effort should be made to solve other types of geophysical problems using extensions, and to solve the explicit deficiencies in the methodology that we identified above.

\section{Conclusion}
Prior information about physical properties of the forward function is important in geophysical inverse problems. We use this information by introducing two specific sample extensions tailored for travel time prediction. These extensions provide a continuous parameter subspace inside which the forward function values are known almost for free given a single travel time calculation, which essentially minimises computational overhead for exploring the parameter space. A deterministic sampling algorithm is proposed to find the most informative samples and their extensions based on a dictionary of possible source-to-receiver ray geometries. We further propose an extensions-based tomography algorithm, which calculates posterior (marginal) distributions analytically using tens of the deterministically chosen samples. We apply the algorithm to a 2D synthetic tomography example, and the results exhibit some consistent features compared to those obtained from a Monte Carlo inversion which uses 1 million random samples and associated forward evaluations. However, certain biases are observed in the inversion results from the proposed algorithm. By the manipulation of individual variables, we pinpoint two major sources of errors that lead to the biases. We outline possible strategies to mitigate these errors. This work indicates that extensions provide valuable information which is currently ignored, yet which might be used to decrease the computational cost of solving nonlinear problems and performing uncertainty analyses.

\section{ACKNOWLEDGMENTS}
We thank the Edinburgh Imaging Project (EIP - \url{https://blogs.ed.ac.uk/imaging/}) sponsors (BP and TotalEnergies) for supporting this research.

\bibliographystyle{plainnat}  
\bibliography{reference}

\appendix

\section{Detailed Inversion Setup}
\label{ap:inversion_setup}

\section{Validation of the on-ray extension}
The on-ray extension defined in the main text is not absolutely valid. When we decrease the slowness value of an on-ray cell, a new ray path could emerge that exhibits a shorter travel time. This situation can arise if the new ray path has a larger ray path length within the considered cell. To illustrate this concept, assume we have two ray paths $R_1$ and $R_2$ that transverse a common cell $m_t$ between a source and a receiver location. Suppose that ray $R_1$ is faster than $R_2$. Say the ray path lengths of $R_1$ and $R_2$ in cell $m_t$ are $l_1$ and $l_2$, respectively. If we decrease the slowness value of $m_t$ by an amount $\Delta m_t$, the travel times along ray $R_1$ and $R_2$ would decease by $l_1\Delta m_t$ and $l_2\Delta m_t$, respectively. When $l_1 < l_2$, the reduction in travel time $l_1\Delta m_t <l_2\Delta m_t$, which might lead to a situation where ray $R_2$ becomes faster than $R_1$. In this case the on-ray extension becomes invalid. 

Nevertheless, the on-ray extension can be valid under certain approximations. Say that the grid cells used to parametrise the slowness field are vanishingly small. Then it is reasonable to assume that every ray has the same path length within every cell it traverses. As a result, the situation described earlier, where a decrease in the on-ray slowness leads to changes in ray path, is less likely to occur, and the on-ray extension remains valid. 

The only thing that can go wrong in applying the on-ray extension in reality is that the grid cells have finite sizes. For this case we can increase the chances that the fastest ray remains after an on-ray slowness decrease by decreasing slowness values in several or many other cells along the ray simultaneously, since this decreases the likelihood that another ray would have a shorter travel time. To explain, consider the above example that contains two rays $R_1$ and $R_2$. When we apply the on-ray extension along $R_1$, we decrease the slowness values for all on-ray cells, rather than only altering the slowness of cell $m_t$. At least one of these cells is an off-ray cell for ray $R_2$ (otherwise $R_1$ and $R_2$ actually represent a same ray). Decreasing slowness of this cell will have no impact on the travel time calculated along $R_2$, but will decrease that calculated along $R_1$. This decreases the probability that $R_2$ would become faster than $R_1$. On the other hand, since we usually apply both the on-ray and off-ray extensions concurrently, the slowness values for the off-ray cells of $R_1$ are also increased. Among these off-ray cells, at least one is by-chance an on-ray cell for $R_2$. Increasing its slowness value also increases the total travel time along $R_2$, further decreasing the chance that $R_2$ becomes faster than $R_1$. In addition, when grid cells are relatively small, the difference in ray path lengths within a same cell (e.g., $|l_1 - l_2|$ within cell $m_t$) tends to be small. This ensures that even in cases where the on-ray extension loses validity (which is already quite unlikely as discussed above), the resulting error in the estimated travel time remains small and potentially negligible. In this context, it is reasonable to assume that the on-ray extension holds, particularly when considering the approximation of small grid cells.
\label{ap:A}

\label{lastpage}
\end{document}